\documentclass[%
 reprint,
 amsmath,amssymb,
 aps,
prb
]{revtex4-2}
\usepackage{graphicx}
\usepackage{dcolumn}
\usepackage{bm}

\begin{document}

\preprint{APS/123-QED}

\title{Quantum Computing for Phonon Scattering Effects on Thermal Conductivity}

\author{Xiangjun Tan$^1$}
 \email{xiangjun.tan@student.unsw.edu.au}
\affiliation{%
$^1$School of Physics, The University of New South Wales, Sydney, New South Wales 2052, Australia
}%

\date{\today}

\begin{abstract}
Recent investigations have demonstrated that multi-phonon scattering processes substantially influence the thermal conductivity of materials, posing significant computational challenges for classical simulations as the complexity of phonon modes escalates. This study examines the potential of quantum simulations to address these challenges, utilizing Noisy Intermediate Scale Quantum era (NISQ) quantum computational capabilities and quantum error mitigation techniques to optimize thermal conductivity calculations. Employing the Variational Quantum Eigensolver (VQE) algorithm, we simulate phonon-phonon contributions based on the Boltzmann Transport Equation (BTE). Our methodology involves mapping multi-phonon scattering systems to fermionic spin operators, necessitating the creation of a customized \textit{ansatz} to balance circuit accuracy and depth. We construct the system within Fock space using bosonic operators and transform the Hamiltonian into the sum of Pauli operators suitable for quantum computation. By addressing the impact of depolarization and non-unitary noise effects, we benchmark the noise influence and implement error mitigation strategies to develop a more efficient model for quantum simulations in the NISQ era.
\end{abstract}

\maketitle

\section{Introduction}
Despite significant advancements in simulating multi-phonon scattering processes, current computational models still struggle with scalability as phonon modes increase. This study introduces a novel quantum simulation approach, anticipated by Feynman five decades ago \cite{cite-key}, optimized for the NISQ era. By developing an effective reduced-complexity model, we aim to significantly cut calculation times and enhance the fidelity of simulations, addressing a critical gap in thermal conductivity research. 

Some physicists have successfully demonstrated that quantum simulation could deploy the system described by the second quantization for High-Energy Physics \cite{doi:10.1126/science.1217069, PRXQuantum.4.027001}. In particular, Baha Balantekin \textit{et,al} from Wisconsin have set up the benchmark of the reduced toy model as part of the dark matter interaction and detection calculation with error mitigation process \cite{PhysRevC.104.024305}. They evaluated the ground state energy for a toy Lipkin model on a real quantum simulator and gave out the general form of the extension to the $N>3$ up to a general case.

Transitioning from high-energy to condensed matter physics, classical-quantum hybrid algorithms have been instrumental in simulating systems where electron-phonon interactions are pivotal \cite{Hensgens_2017, Cervera_Lierta_2018, Geller_2022, Macridin_2018, Li_2020}. For instance, the study in Ref.\cite{Macridin_2018} managed to simplify the electron-phonon coupling system by approximating the phonon term as harmonic and truncating high-frequency contributions. In the NISQ era, simulating the phonon using the lattice framework on a quantum computer seems impossible due to the limited qubit numbers. The phonon is described as collective excitation by more than hundreds of atoms, meaning it needs to occupy too many sites corresponding to the qubit usage. Since some error mitigation strategies have come out and become available \cite{RevModPhys.95.045005, PEC, Takagi2022, Giurgica_Tiron_2020}, it's essential to build a more compatible and efficient model that is suitable for quantum simulation in this time to gain new physics insight. Especially considering the real-world problem with non-unitary noise operators' contributions like depolarization errors and how to mitigate the effect of noise and measurement error.

In this paper, we aim to deploy the quantum simulation to enhance the calculation process for thermal conductivity. The contribution of phonons in thermal conductivity is given out by the BTE, which includes the term Phonon lifetime \text{$\tau_\lambda$} \cite{Feng_2016, Ghosh_2023}. This can be derived from the transition probability by calculating the Fermi Golden Rule as a function determined by the ground state energy like the ShengBTE package \cite{ShengBTE_2014, PhysRevB.86.174307}, which can be simulated by the VQE algorithm or the analogue quantum simulation via quantum platform \cite{Tilly_2022, Daley2023, doi:10.1126/science.ade7651, weckesser2024realizationrydbergdressedextendedbose}. Our challenges are mapping the phonon scattering system to the fermionic spin operator and setting up the effective \textit{ansatz}, which could balance the circuit's accuracy and depth (gate cost). The popular \textit{ansatz} as Unitary Coupled Cluster (UCC) like and N-Local like are not sufficient under this situation \cite{Anand_2022, romero2018strategies, VQE_method}, so we will modify our own \textit{ansatz} to adapt the feature of the phonon scattering system. The second quantized Hamiltonian should be mapped to the form in the Sum of the Pauli product, then converted to the Pauli operator gates to get the expectation value in a quantum instance. However, the mapping process differs from the fermion system mapping to the fermionic operators like quantum gates on qubits. The wavefunction for fermion will be asymmetry, but for boson, it will be symmetry, which means we will construct our system under Fock space with Bosonic number operators \cite{PhysRevLett.131.033604, busnaina2024quantum}. The approximate transformation will map the Hamiltonian into the Pauli operator sequence that could be run on quantum computers. The eigenvalue evaluation in the iteration process will be a fully quantum algorithm, but classical computing will be used to optimise during the iteration steps. However, the evaluation process will not be perfect since the Non-unitary noise operator affects the Pauli gate for Hamiltonian configuration; it will cause the Sum of the Pauli product to become non-hermitian, corresponding to the non-real value eigenvalue for our qubits represented system. The noise model will be customized so we can set up the benchmark to see how the noise will affect our outcome and examine quantum error mitigation strategies \cite{qiskit2024}.

\section{Theory of Thermal Conductivity}
\subsection{Theoretical Model of Thermal Conductivity}
The thermal conductivity is the coefficient between the heat flux \text{J} and the temperature gradient \text{$\nabla T$}.
\begin{equation}
\mathbf{J}=-\kappa \cdot \nabla T
\end{equation}

From the BTE in Ref.\cite{PhysRevB.93.045202, Ghosh_2023, ShengBTE_2014}, the thermal conductivity $\kappa$ for a sample with volume $V$ along the direction \textbf{$i,i\in{x,y,z}$} could be expressed as
\begin{equation}
\kappa_i=\frac{1}{V} \sum_\lambda v_{i, \lambda}^2 c_\lambda \tau_\lambda
\end{equation}

Where $v_i$ is the projection of phonon group velocity on direction \textbf{$i$}, and $c_\lambda$ is the specific heat of phonon per mode, in proper non-cryogentic temperature interval with the frequency modes \text{$\hbar\omega_\lambda$} and the Boltzmann constant \text{$k_B$} given by
\begin{equation}
c_\lambda=k_B\left(\frac{\hbar \omega_\lambda}{k_B T}\right)^2 \frac{e^{ \left(\hbar \omega_\lambda / k_B T\right)}}{\left[e^{ \left(\hbar \omega_\lambda / k_B T\right)}-1\right]^2}
\label{eq:cv}
\end{equation}

The $\tau_\lambda$ is the lifetime that comes from the phonon scattering rate with initial and final states calculated by
\begin{equation}
\begin{aligned}
\tau_\lambda=\frac{1}{\Gamma}=\frac{1}{\frac{2 \pi}{\hbar}\left|\left\langle f\left|\hat{H}\right| i\right\rangle\right|^2 \delta\left(E_i-E_f\right)}
\end{aligned}
\end{equation}
A key to calculating the phonon lifetime is that the first-order phonon vibration and second-order phonon scattering terms will not contribute anything to the thermal conductivity due to they are harmonic and act as constant. The terms that will significantly affect the properties will be the three-phonon scattering process and the four-phonon process will be a higher-order correction for the theoretical values. Within the effective model configuration, the effective lifetime terms we are looking forward to is
\begin{equation}
    \frac{1}{\tau_\lambda}=\frac{1}{\tau_{\lambda_3}}+\frac{1}{\tau_{\lambda_4}}
\end{equation}

\subsection{The Effective Hamiltonian of Phonon Scattering}
With the effective assumption given in the previous part, the ground state energy should be estimated before the calculation process for the Fermi Golden Rule. The system Hamiltonian can be expressed as the sum of harmonic terms and anharmonic terms:
\begin{equation}
\hat{H}=\hat{H}_0+\hat{H}_3+\hat{H}_4+\cdots,
\end{equation}
where the components are defined as follows:

\begin{equation}
\begin{aligned}
\hat{H}_0 &= \sum_\lambda \hbar \omega_\lambda\left(a_\lambda^{\dagger} a_\lambda+\frac{1}{2}\right),
\end{aligned}
\end{equation}

Here we introduced the displacement operator \text{$u_i=(a^{\dagger}_{-\lambda_i}+a_{\lambda_i})$}, with the order \text{n} we can define the scattering matrix element \text{${E_V}_{\lambda\lambda_1...\lambda_{n-1}}^{(n)}$} as
\begin{equation}
    {E_V}_{\lambda\lambda_1...\lambda_{n-1}}^{(n)}=\sum_{bb_1...b_{n-1}}\sum_{aa_1...a_{n-1}}\Phi^{(n)a}_{bb_{1}...b_{n-1}}\frac{e^{iq|r_a|}}{\sqrt{m_{\lambda}...m_{\lambda_{n-1}}}}
\end{equation}

Where the force constant \text{$\Phi^{(n)a_n}_{bb_{1}...b_{n-1}}$} is defined with potential energy \text{$V$}
\begin{equation}
    \Phi^{(n)a_n}_{bb_{1}...b_{n-1}}=\frac{\partial^{n} V}{{\partial r_{a,b}}^{n}}
\end{equation}

The n-th coupling constant $H^{(n)}_{\lambda\lambda_{1}...\lambda_{n-1}}$ with delta \text{$\delta_q$} function and normalizing constant \text{$G$} will be
\begin{equation}
    H^{(n)}_{\lambda\lambda_{1}...\lambda_{n-1}}=G\frac{\Phi^{(n)a_n}_{bb_{1}...b_{n-1}}}{\sqrt{\omega_\lambda\omega_{\lambda_1} \omega_{\lambda_2}...\omega_{\lambda_{n-1}}}}\delta_{q}
\end{equation}

Based on the definition above the $\hat{H}_3$ and $\hat{H}_4$ can be expressed as
\begin{equation}
\begin{aligned}
\hat{H}_3 = \sum_{\lambda \lambda_1 \lambda_2} H_{\lambda \lambda_1...\lambda_2}^{(3)}
u_{\lambda} u_{\lambda_1} u_{\lambda_2}
\label{H3}
\end{aligned}
\end{equation}

\begin{equation}
\begin{aligned}
\hat{H}_4 = \sum_{\lambda \lambda_1 \lambda_2 \lambda_3} H_{\lambda \lambda_1 \lambda_2 \lambda_3}^{(4)}
u_{\lambda} u_{\lambda_1} u_{\lambda_2} u_{\lambda_3}
\end{aligned}
\end{equation}

Only the anharmonic terms will contribute to the thermal conductivity so that the dominant term will be the third term \text{$H_3$} and a perturbation \text{$H_4$} \cite{HAN2022108179, GU2020120165}. If we do not consider the Umklapp scattering (U process \cite{maznev2014demystifying}), Fig.\ref{fig:3rdorder} below shows the corresponding phonon-phonon splitting and combining common process with the Hamiltonian Eq.\ref{H3} above. Note that the occupation number \text{$n_\lambda$} of a specific phonon mode is

\begin{equation}
    n_\lambda=\frac{1}{\exp \left(\frac{\hbar \omega_\lambda}{k_{B} T}\right)-1}
    \label{eq:on}
\end{equation}

\begin{figure}[!htb]
    \centering
        \includegraphics[width=1\linewidth]{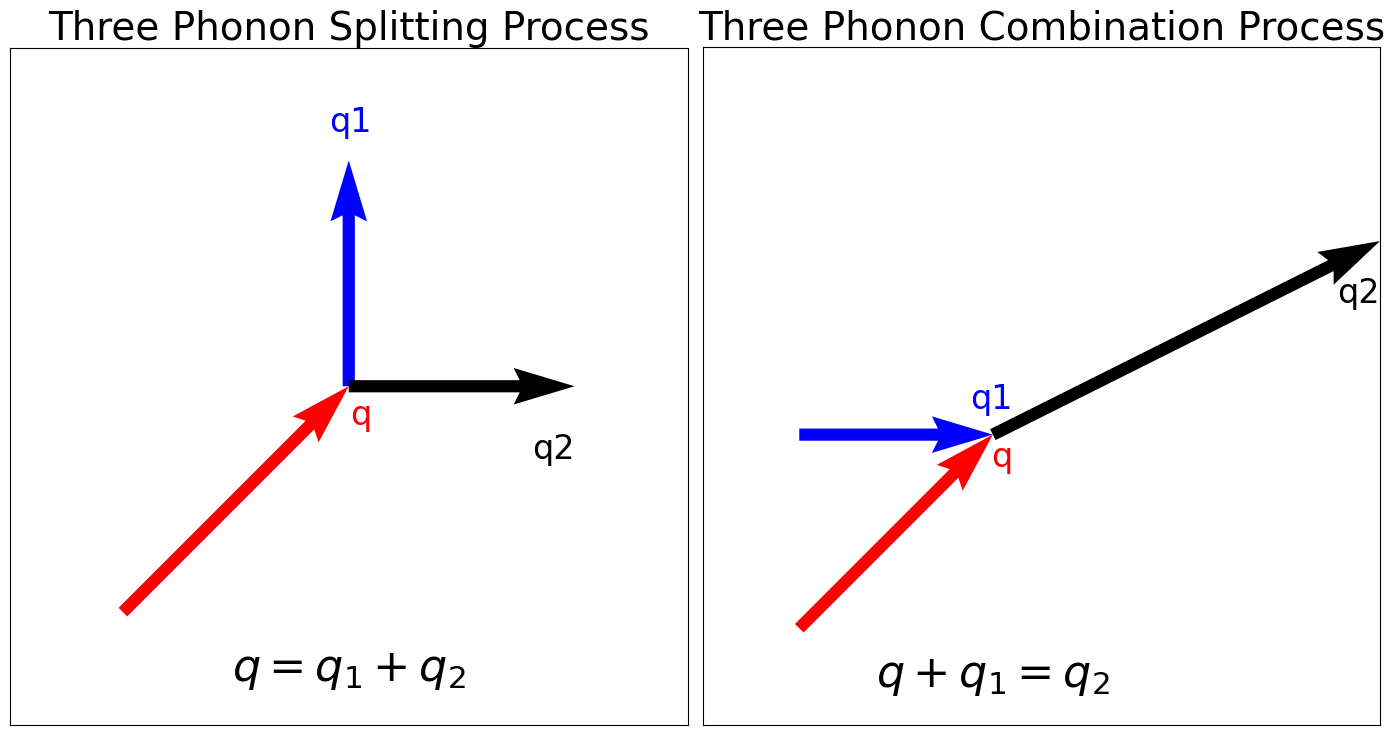}
        \caption{The left figure shows the Normal Process (N process) of phonon with momentum q splitting into two phonons with momentum $q_1$ and $q_2$, and the right figure shows the phonons with momentum q and $q_1$ combining together to become a phonon with momentum $q_2$. For splitting process \text{$|i\rangle=\left|n_\lambda+1, n_{\lambda_1}, n_{\lambda_2}\right\rangle$, $|f\rangle=\left|n_\lambda, n_{\lambda_1}+1, n_{\lambda_2}+1\right\rangle$}, where \text{$\lambda$} is the corresponding phonon wavelength of \text{$q$}.}
        \label{fig:3rdorder}
\end{figure}

\section{Quantum Simulation with VQE algorithm}
\subsection{Mapping the Hamiltonian}

In order to evaluate the contribution of three-phonon and four-phonon scattering cases, only the terms for  $\hat{H}_3+\hat{H}_4$ will be taken into account in the effective Hamiltonian.

\begin{equation}
    \hat{H}_{eff}=\hat{H}_3+\hat{H}_4
\end{equation}

The terms of $\hat{H}_3$ could be expanded to the full terms expression

\begin{equation}
\begin{aligned}
\hat{H}_3 = \sum_{\lambda \lambda_1 \lambda_2} H_{\lambda \lambda_1 \lambda_2}^{(3)}
& \left( a_{-\lambda}^{\dagger} a_{-\lambda_1}^{\dagger} a_{-\lambda_2}^{\dagger} + a_{-\lambda}^{\dagger} a_{-\lambda_1}^{\dagger} a_{\lambda_2} \right. \\
& + a_{-\lambda}^{\dagger} a_{\lambda_1} a_{-\lambda_2}^{\dagger} + a_{-\lambda}^{\dagger} a_{\lambda_1} a_{\lambda_2} \\
& + a_{\lambda} a_{-\lambda_1}^{\dagger} a_{-\lambda_2}^{\dagger} + a_{\lambda} a_{-\lambda_1}^{\dagger} a_{\lambda_2} \\
& \left. + a_{\lambda} a_{\lambda_1} a_{-\lambda_2}^{\dagger} + a_{\lambda} a_{\lambda_1} a_{\lambda_2} \right)
\end{aligned}
\end{equation}\

The bosonic creation operator \( a^\dagger \) in matrix form representation, truncated at \( n_{\text{max}} \) is given by:
\begin{equation}
    a^\dagger_{\text{max}} = \begin{pmatrix}
    0 & 1 & 0 & \cdots & 0 \\
    0 & 0 & \sqrt{2} & \cdots & 0 \\
    0 & 0 & 0 & \ddots & 0 \\
    \vdots & \vdots & \vdots & \ddots & \sqrt{n_{\text{max}}} \\
    0 & 0 & 0 & \cdots & 0
    \end{pmatrix}
\end{equation}

And the bosonic annihilation operator \( a \) in matrix form is the Hermitian conjugate of \( a^\dagger \), given by:
\begin{equation}
    a_{\text{max}} = \begin{pmatrix}
    0 & 0 & 0 & \cdots & 0 \\
    1 & 0 & 0 & \cdots & 0 \\
    0 & \sqrt{2} & 0 & \cdots & 0 \\
    \vdots & \vdots & \ddots & \ddots & \vdots \\
    0 & 0 & \cdots & \sqrt{n_{\text{max}}} & 0
    \end{pmatrix}
\end{equation}

However, this form with second quantization still cannot be directly mapped to the qubit gate sequence unless we represent the Hamiltonian form and convert it into a sum of the Pauli product. Since the particles that take part in the scattering process are all Bosonic, unlike Fermonic operators, which can be converted by the Jordan-Wigner transformation or Bosonic Direct transformation \cite{Somma_2003, Nature_bosonic, the_qiskit_nature_developers_and_contrib_2023_7828768, stavenger2022bosonicqiskit}, the creation operator and annihilation operator must follow by span $U\in SU(2)$ quasi-spin algebra below
\begin{equation}
\left[a_\alpha^{\dagger}, a_\beta^{\dagger}\right]=\left[a_\alpha, a_\beta\right]=0, \quad\left[a_\alpha, a_\beta^{\dagger}\right]=\delta_{\alpha \beta}
\end{equation}
The mapping should set up multi-qubit entanglement,  just consider a case for $m$ phonon $N_n=m$ and $n$ modes per phonon $N_\omega=n$. With the mapping spin basis $|1\rangle=|\uparrow\rangle,|0\rangle=|\downarrow\rangle$, then the number basis could be represented by $\{|0,0,...,0\rangle,|0,1,...,0\rangle,...,|1,1,...,1\rangle\}$, where the physical meaning of these states are given by
\begin{equation}
    \begin{aligned}
        |\text{Vac}\rangle=|0,0,...,0\rangle=|\downarrow\downarrow...\downarrow\rangle\\
        |0\rangle_m=|1,0,...,0\rangle=|\uparrow\downarrow...\downarrow\rangle\\
        |1\rangle_m=|0,1,...,0\rangle=|\downarrow\uparrow...\downarrow\rangle\\
        |\text{$N_\omega$}\rangle_m=|0,0,...,1\rangle=|\downarrow\downarrow...\uparrow\rangle\\
    \end{aligned}
\end{equation}

Here, the subscript $m$ runs over the $m \cdot n$ qubits, indicating the specific phonon mode represented by each qubit. The state $|0\rangle_m$ denotes the phonon mode in its ground state, while $|1\rangle_m$ denotes the mode excited once. Consequently, in this representation, each qubit state corresponds to a distinct configuration of phonon modes across the system, effectively capturing the quantum states of multiple phonons distributed among the $n$ modes. Under this transformation, the matrix dimension of the system will be $2^{m\cdot n}$ by $2^{m\cdot n}$.

So, the number operator could be expressed as
\begin{equation}
n_m=\sum_{n=0}^{N_\omega} n \frac{\sigma_z^{n, m}+1}{2}
\end{equation}

With the definition $|n\rangle_m=\sqrt{n+1}|n+1\rangle_m$, the bosonic creation/annihilation operator mapping to the qubit quasi-spin operator will be
\begin{equation}
\begin{aligned}
a_m^{\dagger}=\sum_{n=0}^{N_\omega-1} \sqrt{n+1} \sigma_{-}^{n, m} 
\sigma_{+}^{n+1, m}\\
a_m=\sum_{n=1}^{N_\omega} \sqrt{n} \sigma_{+}^{n-1, m} \sigma_{-}^{n, m}
\end{aligned}
\end{equation}

Consider a case for three phonon scattering, and each phonon is assigned to one unique frequency. The Hamiltonian terms mapping to the Pauli Operator terms will be

\begin{table}[!htb]
\centering
\begin{tabular}{cc}
\hline
\textbf{Operator} & \textbf{Pauli Operator Representation} \\
\hline
\(a_1^\dagger\) & \(\sigma_{-}^{0,1} \sigma_{+}^{1,1}\) \\
\(a_1\) & \(\sigma_{+}^{0,1} \sigma_{-}^{1,1}\) \\
\(a_2^\dagger\) & \(\sigma_{-}^{0,2} \sigma_{+}^{1,2}\) \\
\(a_2\) & \(\sigma_{+}^{0,2} \sigma_{-}^{1,2}\) \\
\(a_3^\dagger\) & \(\sigma_{-}^{0,3} \sigma_{+}^{1,3}\) \\
\(a_3\) & \(\sigma_{+}^{0,3} \sigma_{-}^{1,3}\) \\
\hline
\end{tabular}
\caption{Mapping of Hamiltonian terms to Pauli operators for a three-phonon system. Here, \(a_m^\dagger\) and \(a_m\) represent the creation and annihilation operators for the m-th phonon, respectively. The superscripts in \(\sigma\) operators, e.g., \(\sigma_{-}^{0,1}\), indicate the action on the phonon state (\(n\)) and phonon index (\(m\)). The first number after the comma in superscript denotes the phonon state, and the second number denotes the phonon mode. Noticed that the coefficient $\sqrt{n}, \sqrt{n+1}$ here is neglected}
\label{tab:hamiltonian_mapping}
\end{table}

\subsection{Hardware Efficient \textit{ansatz} and Toy Model}

According to the Variational Method, the expected ground state energy value is given by:
\begin{equation}
\begin{aligned}
E\left(\theta_1, \cdots, \theta_n\right)&\geq\langle\hat{H}\rangle\\
&=\sum_i \alpha_i\left\langle\psi\left(\theta_1, \cdots, \theta_N\right)\middle|\hat{P}_i\middle| \psi\left(\theta_1, \cdots, \theta_N\right)\right\rangle
\end{aligned}
\end{equation}

Where $\hat{P}_i$ is the Operators mapping by the Effective Hamiltonian within the sum of Pauli product form $\hat{P_i}=\sum_i \hat{O_i}(1)...\hat{O_i}(N)$, and $a_i$ is the scaling normalization coefficient.
\begin{equation}
    \hat{E}\left(\theta_1, \cdots, \theta_n\right)\geq\hat{H}=\sum_i \hat{P}_i=\sum_i \hat{O_i}(1)...\hat{O_i}(N)
\end{equation}

For our system, we use the HEA variational form with total qubit number $N$ as our trial wave function\cite{HEA, Funcke2021dimensional}

\begin{equation}
|\psi(\vec{\theta})\rangle=\left[\prod_{i=1}^N R_{\text {rota }}\left(\theta_i\right) \times U_{\text {entangle}}\right] \times R_{\text {rota }}\left(\theta_{N+1}\right)\left|\psi_{\text {init}}\right\rangle
\end{equation}

Here, in the most common case, we use a Controlled-X gate (CNOT) to perform the entangling operation, which flips the state of the target qubit if the control qubit is in the $|1\rangle$ state. Its matrix representation in the computational basis
\begin{equation}
    U_{\text {entangle}}=\text{CNOT} =
    \begin{pmatrix}
    1 & 0 & 0 & 0 \\
    0 & 1 & 0 & 0 \\
    0 & 0 & 0 & 1 \\
    0 & 0 & 1 & 0 \\
    \end{pmatrix}
\end{equation}

Due to the matrix dimension of simulation, which will significantly increase the performance cost, our toy model will build up with three phonon cases with three modes \text{$\omega_\lambda=[0.5,1.0]$} per phonon to satisfy the relationship of combining and splitting process \text{$|q|=1.0,|q1|=|q2|=0.5$}, and the rest of the phonon modes will be truncated. The qubit usage $N$ for three-phonon scattering will be

\begin{equation}
    N=N_n\cdot N_\omega =3\cdot2=6
\end{equation}

For multi-qubit entanglement, the total state will be the Kronecker product by all of the qubit states $|\psi\rangle=|\psi_1\rangle\otimes|\psi_2\rangle\otimes...\otimes|\psi_N\rangle$. The corresponding quantum circuit is shown below, with a complete gate set that aims to sweep all the possible states by iterating the parameters to avoid the local minimum.
\begin{figure}[!htb]
    \centering 
        \includegraphics[width=1\linewidth]{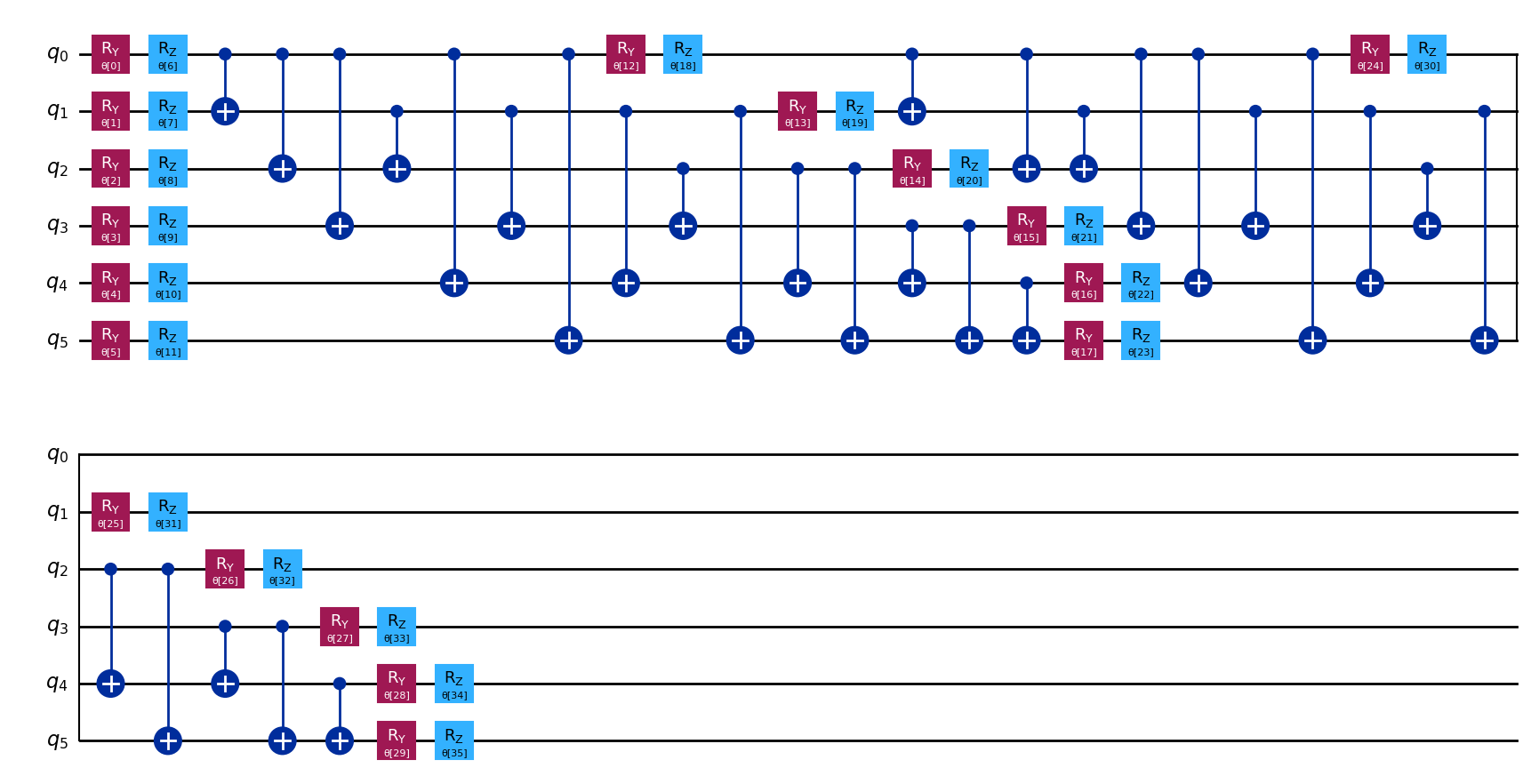}
        \caption{The $6$ qubits whole entanglement quantum circuit of parameterised Hardware Effective  \textit{Ansatz} with $2$ repeats has $36$ parameters. Each qubit will align to six parameter sets of $R_{X/Y/Z}$, and $30$ CNOT gates will linearly entangle all qubits.
        }
        \label{fig:Ansatz}
\end{figure}

However, this will be noisy due to its complexity when we evaluate the energy. To reduce the circuit layer and the number of total gates, we have designed a more efficient \textit{ansatz} with less \text{$R_{Y/Z}$} operation and also remove the extra CNOT gate by just using one CNOT to operate the last and the first qubit. To address the complexity and noise in evaluating the energy due to the feature of our real amplitude system, we exclusively use \text{$R_{Y}$} gates. 

\begin{figure}[!htb]
    \centering 
        \includegraphics[width=1\linewidth]{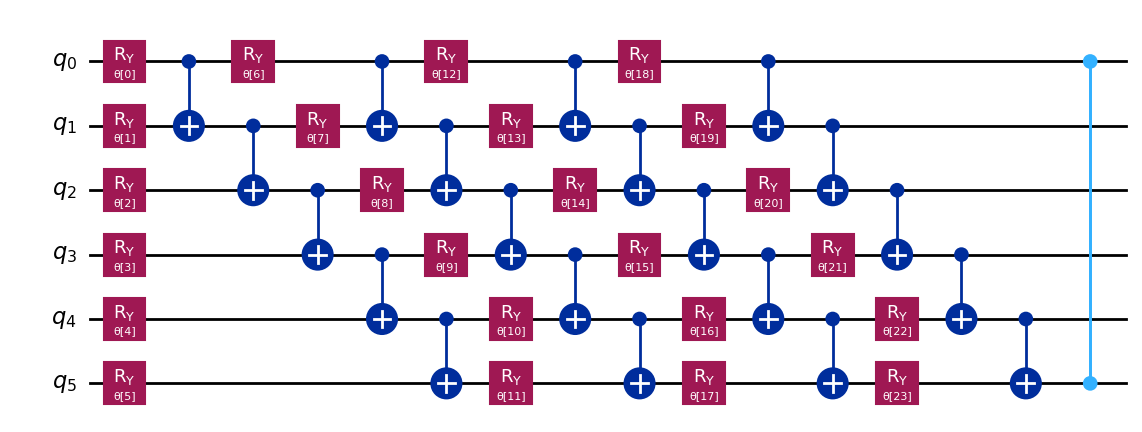}
        \caption{The customized effective \textit{ansatz} significantly reduces gate and circuit layer usage; only $24$ parameters are assigned to the single qubit rotation \text{$R_{Y}$} gate, $20$ CNOT gate, and $1$ CZ gate (light blue one) to create the entanglement.
        }
        \label{fig:Ansatz_custom}
\end{figure}

According to the Fermi Gorden Rule, the transition probability for three phonon splitting interactions \text{$\lambda\rightarrow\lambda_1+\lambda_2$} with occupation number \text{$n_\lambda$} defined in Eq.\ref{eq:on} could be written as
\begin{equation}
\begin{aligned}
\Gamma=\frac{2 \pi}{\hbar}\left|\left\langle f\left|\hat{H}_3\right| i\right\rangle\right|^2 \delta\left(E_i-E_f\right)\\\sim n_\lambda\left(1+n_{\lambda_1}\right)\left(1+n_{\lambda_2}\right)\left|H_{\lambda \lambda_1 \lambda_2}^{(3)}\right|^2
\end{aligned}
\label{eq: gamma}
\end{equation}

For the splitting process shown in Fig.\ref{fig:3rdorder} we have
\begin{equation}
    \langle f|\hat{H}_3|i\rangle=\langle 0|(a_{-\lambda} a_{\lambda_1}^{\dagger} a_{\lambda_2}^{\dagger})^{\dagger} \hat{H}_3 (a_{-\lambda}^{\dagger} a_{\lambda_1} a_{\lambda_2})|0\rangle
    \label{eq:fgr}
\end{equation}

With the relation \text{$a^{\dagger}_{\lambda}|n_\lambda\rangle=\sqrt{n_\lambda+1}|n_\lambda+1\rangle$} and \text{$a_{\lambda}|n_{\lambda}\rangle=\sqrt{n_{\lambda}}|n_{\lambda}-1\rangle$}

\begin{equation}
    =n_\lambda\left(1+n_{\lambda_1}\right)\left(1+n_{\lambda_2}\right)\langle (a_{-\lambda} a_{\lambda_1}^{\dagger} a_{\lambda_2}^{\dagger})^{\dagger} \hat{H}_3 (a_{-\lambda}^{\dagger} a_{\lambda_1} a_{\lambda_2})\rangle
\end{equation}

The second expression in Eq.\ref{eq: gamma} shows the classical computing method by using DFT calculation \cite{ShengBTE_2014}, here in our hybrid algorithm, we will estimate the Eq.\ref{eq:fgr} directly. In our toy model, there are $6$ qubits in use and mapping \text{$2$} modes per phonon with group velocity \text{$8500$m/s} as baseline. The temperature we take for \text{$100$K to $300$K} so the Eq.\ref{eq:cv} can be used to calculate the specific heat, the normalizing constant \text{$G$} will be set up with \text{$1$} to reduce the complexity, note that all allowed process need to satisfy the energy conservation that \text{$E_\lambda=E_{\lambda_1}+E_{\lambda_2}$}.

\subsection{Wick Expansion Method and Two-point Correlation Functions}

In some materials, the four-phonon scattering will significantly reduce the thermal conductivity \cite{PhysRevB.96.161201, PhysRevB.101.125416}. By the expansion of the \text{$\hat{H}_4$}, the full terms form of the four phonon scattering part will be
\begin{equation}
\begin{aligned}
\hat{H}_4 &= \sum_{\lambda \lambda_1 \lambda_2 \lambda_3} H_{\lambda \lambda_1 \lambda_2 \lambda_3}^{(4)} \\
&\quad \times \left( a_{-\lambda}^{\dagger} a_{-\lambda_1}^{\dagger} a_{-\lambda_2}^{\dagger} a_{-\lambda_3}^{\dagger} + a_{-\lambda}^{\dagger} a_{-\lambda_1}^{\dagger} a_{-\lambda_2}^{\dagger} a_{\lambda_3} \right. \\
&\quad + a_{-\lambda}^{\dagger} a_{-\lambda_1}^{\dagger} a_{\lambda_2} a_{-\lambda_3}^{\dagger} + a_{-\lambda}^{\dagger} a_{-\lambda_1}^{\dagger} a_{\lambda_2} a_{\lambda_3} \\
&\quad + a_{-\lambda}^{\dagger} a_{\lambda_1} a_{-\lambda_2}^{\dagger} a_{-\lambda_3}^{\dagger} + a_{-\lambda}^{\dagger} a_{\lambda_1} a_{-\lambda_2}^{\dagger} a_{\lambda_3} \\
&\quad + a_{-\lambda}^{\dagger} a_{\lambda_1} a_{\lambda_2} a_{-\lambda_3}^{\dagger} + a_{-\lambda}^{\dagger} a_{\lambda_1} a_{\lambda_2} a_{\lambda_3} \\
&\quad + a_{\lambda} a_{-\lambda_1}^{\dagger} a_{-\lambda_2}^{\dagger} a_{-\lambda_3}^{\dagger} + a_{\lambda} a_{-\lambda_1}^{\dagger} a_{-\lambda_2}^{\dagger} a_{\lambda_3} \\
&\quad + a_{\lambda} a_{-\lambda_1}^{\dagger} a_{\lambda_2} a_{-\lambda_3}^{\dagger} + a_{\lambda} a_{-\lambda_1}^{\dagger} a_{\lambda_2} a_{\lambda_3} \\
&\quad + a_{\lambda} a_{\lambda_1} a_{-\lambda_2}^{\dagger} a_{-\lambda_3}^{\dagger} + a_{\lambda} a_{\lambda_1} a_{-\lambda_2}^{\dagger} a_{\lambda_3} \\
&\quad \left. + a_{\lambda} a_{\lambda_1} a_{\lambda_2} a_{-\lambda_3}^{\dagger} + a_{\lambda} a_{\lambda_1} a_{\lambda_2} a_{\lambda_3} \right).
\end{aligned}
\end{equation}

In order to reduce the complexity and the gate cost to attach the minimum number of qubits usage, the Wick theorem with operators \text{$A_i$} applied as follows:
\begin{equation}
\begin{aligned}
&\left\langle 0\middle|A_1 A_2 A_3 A_4\middle| 0\right\rangle \\
&= \left\langle 0\middle|A_1 A_2\middle| 0\right\rangle\left\langle 0\middle|A_3 A_4\middle| 0\right\rangle\\
&\quad+\left\langle 0\middle|A_1 A_3\middle| 0\right\rangle\left\langle 0\middle|A_2 A_4\middle| 0\right\rangle\\
&\quad+\left\langle 0\middle|A_1 A_4\middle| 0\right\rangle\left\langle 0\middle|A_2 A_3\middle| 0\right\rangle,
\end{aligned}
\end{equation}  

So in the calculation process of the Fermi Golden Rule, the four phonon interaction operators could be decomposed by applying Wick's theorem, the four-phonon interaction Hamiltonian can be decomposed into a sum of two-body terms:
\begin{equation}
\begin{aligned}
&\langle f|\hat{H}_4|i \rangle = \\
&\quad \sum_{\lambda \lambda_1 \lambda_2 \lambda_3} H_{\lambda \lambda_1 \lambda_2 \lambda_3}^{(4)} \Big( \langle f|a_{-\lambda}^{\dagger} a_{\lambda}|i \rangle\langle f|a_{-\lambda_1}^{\dagger} a_{\lambda_2}|i \rangle \\
&\hspace{1em} + \langle f|a_{-\lambda}^{\dagger} a_{\lambda_1}|i \rangle\langle f|a_{-\lambda_2}^{\dagger} a_{\lambda_3}|i \rangle \\
&\hspace{1em} + \langle f|a_{-\lambda}^{\dagger} a_{\lambda_3}|i \rangle\langle f|a_{-\lambda_1}^{\dagger} a_{\lambda_2}|i \rangle \Big)\\
&\quad + \text{...other two-body terms}
\end{aligned}
\label{eq:four_phonon_decomposition}
\end{equation}

This approximation simplifies the computation of the Fermi Golden Rule by reducing the four-phonon interaction to a sum of products of two-point correlation functions \cite{Lepori_2023}.

\section{Result}

\subsection{Result on noiseless quantum simulator}

\begin{figure}[htb!]
    \centering
    \includegraphics[width=1\linewidth]{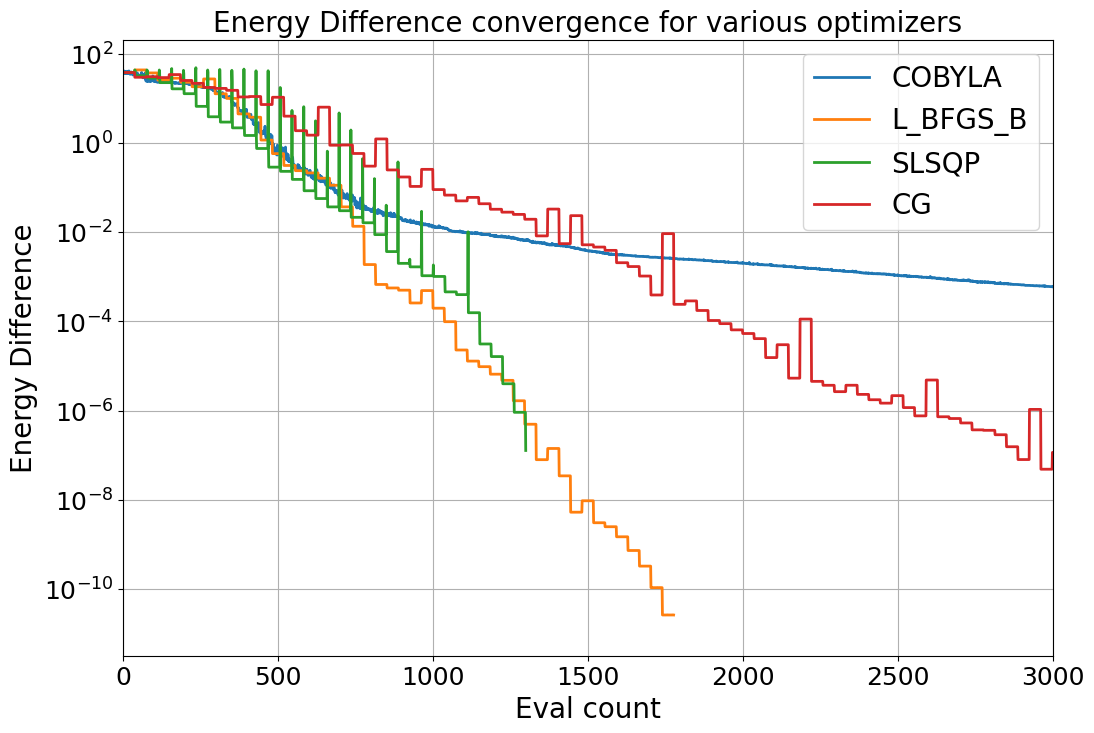}
    \caption{The energy convergence diagram with different optimizers on the noiseless simulator. According to the different types of optimizers, it is certain that the energy difference will be lower than \text{$1e-3$}.}
    \label{fig:output1}
\end{figure}

The different optimizers in Scipy have been used to determine how the parameters will be iterated. For the noiseless case, we set the shots with \text{$4096$} times per iteration, and we could see that the \text{L\_BFGS\_B} will be the fastest gradient-free optimizer. The EffcientSU(2) \textit{ansatz} constructed in Fig.\ref{fig:Ansatz} works well after the transpilation process in all of the cases with energy differences less than $0.001$ with the relative error $0.02\%$ shown in Fig.\ref{fig:output1}.

\begin{figure}[htb!]
    \centering
    \includegraphics[width=1\linewidth]{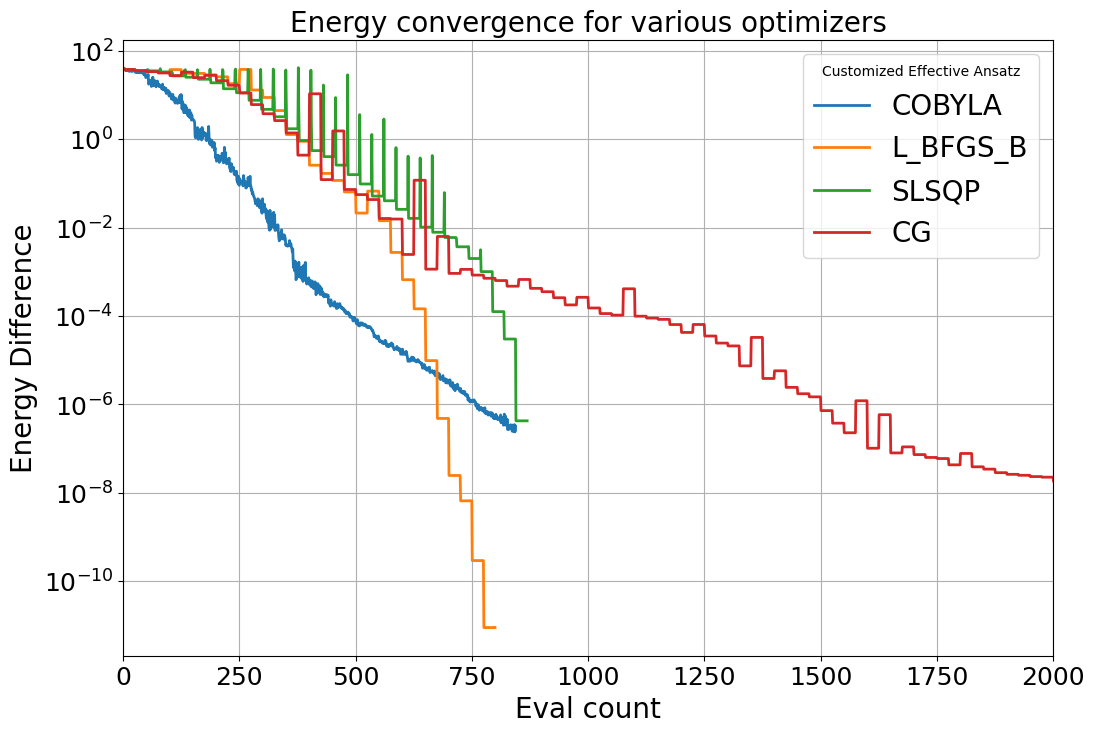}
    \caption{The energy convergence diagram with different optimizers on the noiseless simulator for the customized effective \textit{ansatz}. In comparison, the optimizer like \text{L\_BFGS\_B} only iterates $760$ times then convergence, which is half of the EffcientSU(2) \textit{ansatz} and other optimizers results in all tend to an error within \text{1e-6}.}
    \label{fig:output_custom}
\end{figure}

For the customized effective \textit{ansatz}, which employs fewer gates, faster convergence is observed, and the energy difference is suppressed to a lower level as shown in Fig.\ref{fig:output_custom}. This \textit{ansatz} is superior to the EfficientSU(2) \textit{ansatz}, providing better performance with less gate usage than we expected.

However, the measurement outcome will not be accurate due to the imperfect gate fidelity \cite{RevModPhys.95.045005}. We will deploy and evaluate the benchmark using different error mitigation strategies based on the limitations of existing Quantum Simulators in the NISQ era. Noise in quantum systems typically results in non-unitary evolution, representing processes that are not reversible.  In order to compare and investigate the mitigation ability, all of the single qubit gates will assume $100\%$ fidelity. The noise will only influence the Echoed Cross-Resonance (ECR) gate and measurements. Note that in our algorithm, ECR gate will be the operation to make the qubits entanglement due to the current support devices. Unlike the CNOT gate, the matrix form of the ECR gate by BIG endian convention is
\begin{equation}
ECR_{q_0, q_1}=\frac{1}{\sqrt{2}}\left(\begin{pmatrix}
0 & 1 & 0 & i \\
1 & 0 & -i & 0 \\
0 & i & 0 & 1 \\
-i & 0 & 1 & 0
\end{pmatrix}\right)=\frac{1}{\sqrt{2}}(I X-X Y)
\end{equation}

Due to all of the ansatz creation being based on the IBM quantum basic gate sets, we have to convert any circuits that include a CNOT gate into an ECR gate set. Below in Fig.\ref{fig:CNOT} is an example of such a conversion, demonstrating the equivalence in quantum approximate operations between the CNOT and ECR gates with single-qubit rotations.
\begin{figure}[!htb]
    \centering
    \includegraphics[width=1
    \linewidth]{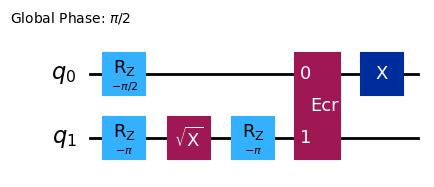}
    \caption{Conversion of the CNOT gate into an ECR gate with surrounding rotations. The ECR gate here replaces a standard CNOT with additional single-qubit $R_Z$ rotations to ensure equivalent functionality. The circuit features a global phase of $\pi/2$, vital for maintaining the coherence of quantum states throughout the transformation.}
    \label{fig:CNOT}
\end{figure}

We have built the noise model on our local simulator to investigate the influence of multi-qubit operation, which will be valuable in estimating the fidelity requirements. Considering the depolarization error will be the dominant term of operation on \text{n} qubits, with the definition
\begin{equation}
\mathcal{E}(\rho)=\frac{p I}{2^n} \operatorname{Tr}(\rho)+(1-p) \rho
\end{equation}

Here is an expression to estimate the ECR fidelity without a Randomized benchmarking, where \text{$I$} is the identity matrix, \text{$\rho$} is the density matrix. So the exact operation after applying error and the fidelity for the ECR gate (\text{$n=2$}) with only depolarization error can be defined as

\begin{equation}
\tilde{U}_{\mathrm{ECR}}(\rho)=\sum_{k=0}^{15} E_k U_{\mathrm{ECR}} \rho U_{\mathrm{ECR}}^{\dagger} E_k^{\dagger}
\end{equation}

\begin{equation}
F\left(U_{\mathrm{ECR}}, \tilde{U}_{\mathrm{ECR}}\right)=\frac{1}{16}\left|\operatorname{Tr}\left(U_{\mathrm{ECR}}^{\dagger} \tilde{U}_{\mathrm{ECR}}\right)\right|^2
\end{equation}

The relation between the gate imperfect gate fidelity affected by the depolarization error and the estimated boundary can be explored with our two types \textit{ansatz}.
\begin{table}[!htb]
\centering
\begin{tabular}{lcr}
\hline
ECR Gate Fidelity & Lower bound ($\%$) & Upper bound ($\%$) \\
\hline
\multicolumn{3}{c}{EfficientSU(2) \textit{ansatz}} \\
100 (perfect case) & 99.547 & 100.004 \\
99.999 & 98.974 & 99.724 \\
99.99 & 96.181 & 99.711 \\
99.9 & 93.082 & 94.133 \\
99 & 80.818 & 81.111 \\
98 (worst case) & 60.077 & 61.342 \\
\hline
\multicolumn{3}{c}{Customized effective \textit{ansatz}} \\
100 (perfect case) & 99.634 & 100.004 \\
99.999 & 99.326 & 100.001 \\
99.99 & 99.181 & 99.711 \\
99.9 & 98.318 & 98.877 \\
99 & 85.584 & 86.171 \\
98 (worst case) & 73.538 & 74.371 \\
\hline
\end{tabular}
\caption{The ECR Gate Fidelity with Depolarization error tested with EfficientSU(2) for $2$ repeats and customized effective \textit{ansatz}, the values are compared to the baseline by classical reference level, so the units are represented as percentages. Due to the energy being negative so when noise increases the factor of \text{$\frac{E_{est}}{E_{ref}}$} will reduced.}
\label{tab:ecr_depolarization}
\end{table}

According to the results in Tab.\ref{tab:ecr_depolarization}, we could estimate that for total fidelity to attach \text{$\operatorname{P}(\mu-1 \sigma \leq X \leq \mu+1 \sigma)$}, the \text{$68.27\%$} requirement for ECR gate fidelity with the system using full entanglement EffcientSU(2) \textit{ansatz} with $m$ phonon and $n$ modes per phonon in $k$ repeats will be
\begin{equation}
    F=^{\frac{(m \times n) \times((m \times n)-1)}{2}\cdot k}\sqrt{68.27\%}
\end{equation}

If we can assume all the single qubits are perfectly defined, our toy model will acquire \text{$\frac{(3 \times 2) \times((3 \times 2)-1)}{2}\cdot 2 = 30$} ECR gate, if only ECR error occurred in the result, the final outcome within \text{$1\sigma$} and \text{$3\sigma$} will be 
\begin{equation}
    F_{1\sigma}=^{30}\sqrt{68.27\%}\approx 0.9874,
    F_{3\sigma}=^{30}\sqrt{99.73\%}\approx 0.9999
    \label{eq:ecr}
\end{equation}

For our customized effective \textit{ansatz}, the required fidelity figures can be significantly reduced:
\begin{equation}
    F_{1\sigma}=^{21}\sqrt{68.27\%}\approx 0.9819,
    F_{3\sigma}=^{21}\sqrt{99.73\%}\approx 0.9998
    \label{eq:ecr_cust}
\end{equation}

Compared with Tab.\ref{tab:ecr_depolarization}, our simulation results on the local simulator are consistent with the fidelity estimation shown in Eq.\ref{eq:ecr} and Eq.\ref{eq:ecr_cust} with only the ECR gate error. The result reveals that our customized \textit{ansatz} exhibits enhanced resilience to depolarization noise compared to EfficientSU(2). Unfortunately, the devices we tested at present only have the ECR fidelity with \text{$0.9914$}; in order to get the result in an acceptable range, we have to utilize some quantum error suppression and mitigation strategies \cite{RevModPhys.95.045005}.

\subsection{Quantum Error Mitigation}

We tested with the noisy natural quantum resource since the noiseless case shows the algorithm works well. For the configuration in the mitigated case, we set the optimization level to $1$, enabling gate twirling/dynamical decoupling and ZNE with readout error mitigation. 

\begin{figure}[!htb]
    \centering 
        \includegraphics[width=1
        \columnwidth]{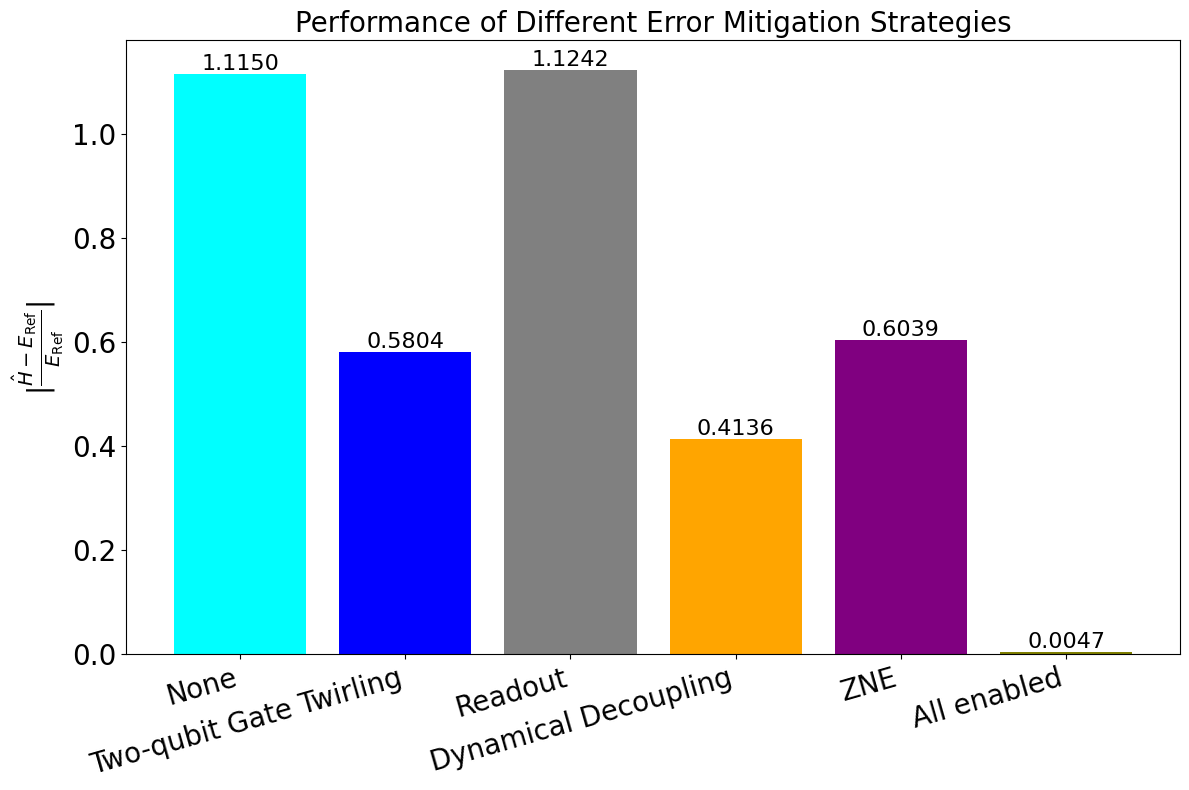}
        \caption{The figure shows the impact of different error mitigation techniques on the computed energy. The standard derivation for those cases are $0.1294, 0.1228, 0.1446, 0.1155, 2.2672, 0.5524$.}
        \label{fig:mitigation}
\end{figure}

As we expected, when enabling all the strategies for quantum error mitigation, the relative measurement outcome has an acceptable range with less than \text{$1\%$}. However, it's important to note that the additional processing required for these techniques can introduce extra variance in the measurements. Interestingly, enabling two-qubit gate twirling and dynamical decoupling significantly reduced the relative error compared to the unmitigated case, with values of $0.580$ and $0.414$ respectively compared to the reference value, and relatively low standard deviations of $0.123$ and $0.116$. This indicates that these techniques effectively reduce errors while maintaining consistency.

On the other hand, readout error mitigation alone showed a higher relative error of $1.124$ with a standard deviation of $0.145$, suggesting that readout errors are substantial and may not be sufficiently mitigated by this strategy alone. The ZNE technique also performed well with a normalized value of $0.604$ and a standard deviation of $2.267$, highlighting its utility in mitigating errors and pointing to a higher variance.

The combined use of all error mitigation techniques yielded the best result, with a normalized value of $0.005$ and a standard deviation of $0.552$, demonstrating the powerful cumulative effect of multiple error mitigation strategies.

\section{Discussion}

\subsection{Error Characterization}
Accurate characterization of qubit performance and error rates is essential for designing efficient quantum circuits and implementing effective error mitigation strategies \cite{e24101467}. The table below provides the average values of these parameters. The relaxation time (\(T_1\)) of the qubits is measured to be 224.67 \(\mu\)s, indicating the average time a qubit remains in its excited state before relaxing to the ground state. The dephasing time (\(T_2\)) is 140.09 \(\mu\)s, which reflects the timescale over which a qubit maintains its coherence, accounting for both energy relaxation and dephasing due to environmental interactions. The qubit frequency is approximately 4.906 GHz, a parameter that influences the qubit's operational speed and its interaction with control pulses. The Error Per Logical Gate (EPLG) is reported to be 2.1\%, providing an estimate of the average error rate per layer quantum gate operation.

\begin{table}[!htb]
\centering
\begin{tabular}{lcr}
    \hline \hline 
    \multicolumn{2}{c}{\textbf{ibm\_brisbane}} \\
    \cline{1-2}
    \textbf{Parameter} & \textbf{Average} \\
    \hline
    $T_1(\mu \mathrm{s})$ & 224.67 \\
    $T_2(\mu \mathrm{s})$ & 140.09 \\
    Qubit frequency (GHz) & 4.906 \\
    EPLG & 2.1 \\
    Pauli-X gate error & 0.002457 \\
    ECR gate error & 0.008471 \\
    Readout error $P(0 \mid 1)$ & 0.0148 \\
    Readout error $P(1 \mid 0)$ & 0.0108 \\
    \hline \hline
\end{tabular}
\caption{Regarding specific gate errors, the Pauli-X gate error is recorded at 0.002457, and the ECR gate error is 0.008471; note that this device only uses ECR operation to get the entanglement without the CNOT gate. The error probability of reading a 0 when the actual state is 1 (\text{$P(0\mid1)$}) is 0.0148, and the probability of reading a 1 when the actual state is 0 (\text{$P(1\mid0)$}) is 0.0108.}
\label{Qubits Error Characterization}
\end{table}

As outlined above, the precise characterization of qubit performance and error rates is crucial for designing efficient quantum circuits and selecting appropriate error mitigation strategies, and we will choose the qubits with the best performance to map because the usage will not fill all the qubits for this device. These parameters form the basis of our subsequent benchmarking tests, where we systematically evaluate the performance of quantum algorithms under realistic operational conditions. What's more, the errors will not fixed all the time, and they will change with time and become larger by the effect of the time accumulations \cite{Takagi2022}.

We chose the following qubits layout in Tab.\ref{table:qubit_mapping} with the best performance for the transformation and compilation process and the rest as the ancilla qubits.
\begin{table}[htb!]
\centering
\begin{tabular}{ccc}
\toprule
\textbf{Physical Qubit} & \textbf{Ansatz Qubit} & \textbf{Classical Register} \\
4 & 1 &  1\\
15 & 2 & 2\\
5 & 0 & 0\\
22 & 3 & 3\\
21 & 4 & 4\\
20 & 5 & 5\\
\hline
\end{tabular}
\caption{Table for \textit{ansatz} logical to device physical qubits mapping, although those logical Qubit on the devices are not directly connected, the rest of the ancillary qubits will play a role as the swapping target.}
\label{table:qubit_mapping}
\end{table}

\subsection{Benchmark}

\begin{figure}[!htb]
        \centering \includegraphics[width=1
        \columnwidth]{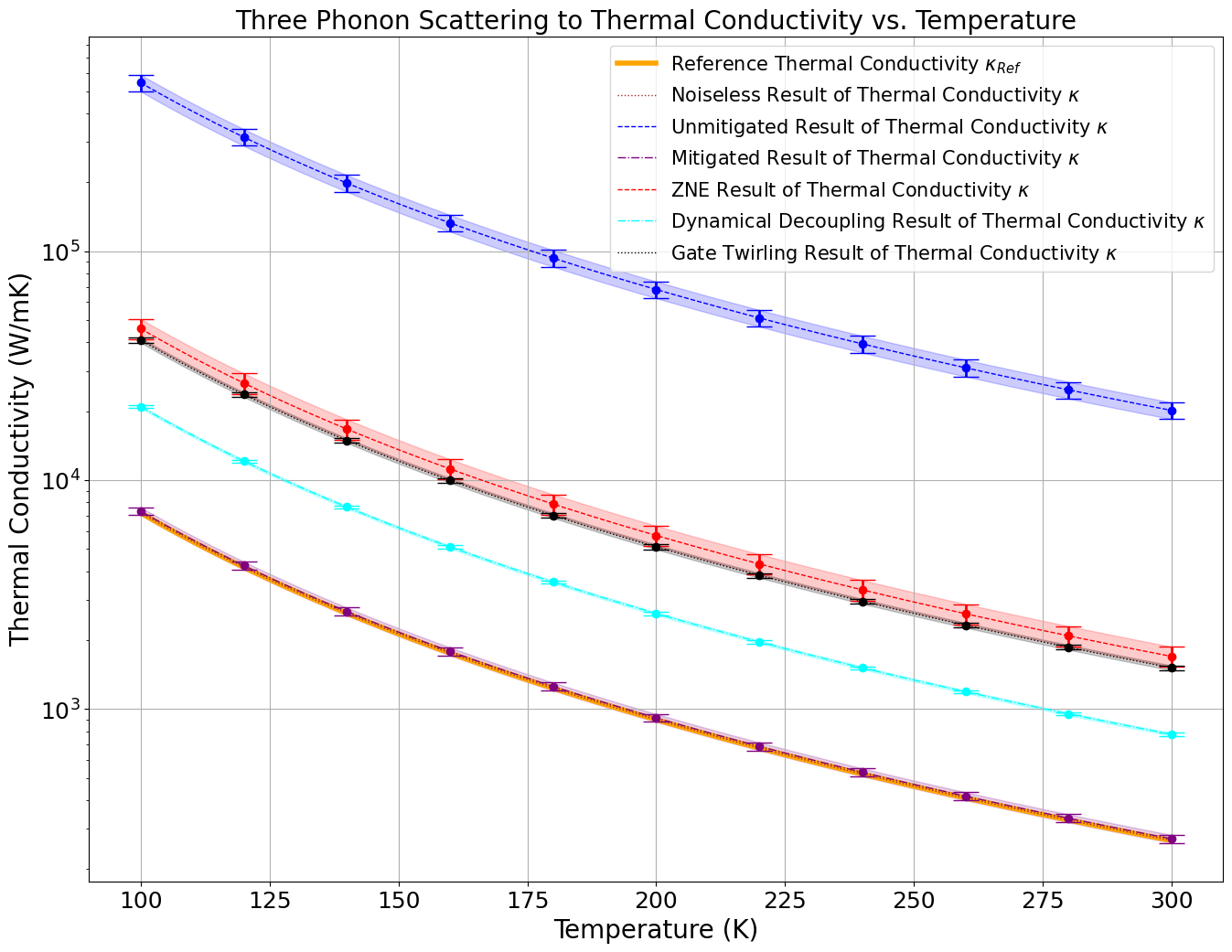}
        \caption{Figure about thermal conductivity results with confidence interval. Values in the y-axis are on a log scale and will not be relevant to real-world materials. The deviation from the unmitigated and reference results highlights the significant impact of quantum noise and other error sources on computational accuracy. The various error-mitigated results greatly benefit from strategies based on our configuration. The best configuration, represented by the purple line, follows the ideal results closely and is shown to be the optimal case.
        }
        \label{fig:thermal}
\end{figure}

Then in this case, the phonon lifetime is \text{$\tau_3=\frac{1}{\Gamma}$} only including the third-order contribution, the fig.\ref{fig:thermal} displays the thermal conductivity results as a function of temperature, showing different computational scenarios: noiseless, unmitigated, and mitigated results compared against a reference thermal conductivity of our toy model evaluated by the customized effective \textit{ansatz} as shown in Fig.\ref{fig:Ansatz_custom}, we have checked the noiseless case had an RMSE value less than \text{$0.01$}.

The orange line in Fig.\ref{fig:thermal} represents the reference level from classical computing, which serves as the benchmark for evaluating the performance of the quantum computational models. The shaded area around the mitigated results visually represents the uncertainty associated with these measurements, although the estimated value is still higher than the ideal case. The width of the interval suggests variability in the mitigation's effectiveness, which means the mitigation strategies will increase the uncertainty of the result.

The benchmark subsection also contains a table that provides a detailed comparison of thermal properties across a range of temperatures from 100 K to 300 K, focusing on specific heat and thermal conductivity under different simulation configurations with customized effective \textit{ansatz} as shown in Tab.\ref{tab: comparison}.
\begin{table}[!htb]
\begin{ruledtabular}
\begin{tabular}{lccccr}
T &\text{$c_v$}\footnote{Note a. Specific Heat(J/K)$\cdot 1e-23$, we should ignore the scale due to this toy model configuration}&\text{$\kappa$}\footnote{Note b. Ideal Thermal Conductivity(W/mK)}&\ Noiseless \text{$\kappa$}&\ Unmitigated&\ Mitigated\\
\hline
100 & 4.1396 & 7200.133 & 7200.134 & 544617.644 & 7327.802 \\
150 & 4.1409 & 2133.613 & 2133.610 & 161386.338 & 2171.445 \\
200 & 4.1414 & 900.153 & 900.154 & 68087.542 & 916.114 \\
250 & 4.1416 & 460.887 & 460.889 & 34861.457 & 469.059 \\
300 & 4.1417 & 266.720 & 266.720 & 20174.654 & 271.449 \\
\end{tabular}
\end{ruledtabular}
\caption{The large discrepancies between the unmitigated, ideal, and noiseless results illustrate the substantial effect of noise in quantum simulations. This highlights the critical importance of error mitigation techniques in maintaining fidelity to expected outcomes. The precise match between mitigated and noiseless results across the table confirms the effectiveness of the applied mitigation strategies. Notably, mitigation can recover the simulation's fidelity to a level indistinguishable from an ideal noiseless scenario.
}
\label{tab: comparison}
\end{table}

Particularly, at lower temperatures ($100$ K to $200$ K), the thermal conductivity values between these scenarios show minimal discrepancies, suggesting that our mitigation strategies are most effective in this range. Conversely, the stark deviation in the unmitigated results highlights the substantial degradation of computational fidelity due to uncorrected quantum errors, with discrepancies as large as over 7000 times the noiseless value at 100 K. This variation emphasizes the necessity of robust error mitigation, as even slight inaccuracies in handling quantum noise can lead to significant errors in computed thermal properties. Moreover, the shaded areas around the mitigated results represent the uncertainty associated with these measurements. The variability in the width of these intervals at different temperatures indicates the fluctuating effectiveness of our mitigation techniques, suggesting that while beneficial, these strategies introduce their own complexities to the accuracy and reliability of quantum simulations. This analysis not only validates the implemented mitigation approaches but also highlights areas for further refinement, particularly in optimizing these strategies to minimize result variability across all tested temperatures.

The model used in these simulations simplifies the complexity of actual materials by reducing the number of quantum states considered by several orders of magnitude and neglecting some real coefficients as $1$. This simplification is necessary for current quantum computational capabilities but does imply certain limitations in directly translating these results to practical applications.

\section{Summary}

This paper explores the significant impact of phonon scattering on thermal conductivity and other physical phenomena through a novel application of quantum computing techniques. We have developed and verified a toy model to simulate three-phonon scattering processes by employing a classical-quantum hybrid algorithm within a quantum computing framework. While this model offers preliminary insights, it is acknowledged that real-world complexities are substantially greater.

Our findings indicate that the EPLG and gate fidelity currently exceed acceptable thresholds for the toy model. However, we have demonstrated the feasibility of achieving results within acceptable limits through targeted circuit optimization and implementing advanced quantum error mitigation strategies. This study primarily utilized bosonic operator mappings focusing on direct mappings to optimize performance. We found that alternative mapping strategies, such as binary mapping, could significantly reduce qubit requirements—potentially halving the number required from six to three in our toy model.

Future research will aim to reduce gate costs and qubit consumption via a more effective ansatz and Hamiltonian mapping. Such advancements are crucial for enhancing the practicality and efficiency of quantum computations for broader applications, not only about thermal conductivity but also including the dynamics of chiral phonons and topological properties \cite{PhysRevResearch.4.L012024, PhysRevB.106.224304, Murta_2020}. Techniques such as the Wick expansion, which reduces many-body interactions to a sum of two-body terms, are promising for minimizing errors associated with multi-qubit gates. Additionally, the challenge of increased uncertainty in measurements, especially in complex scenarios like four-phonon scattering where the frequency per phonon also increases, will be a focal point of continuing investigations.

\begin{acknowledgments}

We express our sincere gratitude to Tiantian Zhang, Shuai Zhang from the Institute of Theoretical Physics, CAS, and Zhanning Wang from UNSW for their invaluable discussions and insights.
\end{acknowledgments}

\bibliography{apssamp}

\begin{thebibliography}{44}%
\makeatletter
\providecommand \@ifxundefined [1]{%
 \@ifx{#1\undefined}
}%
\providecommand \@ifnum [1]{%
 \ifnum #1\expandafter \@firstoftwo
 \else \expandafter \@secondoftwo
 \fi
}%
\providecommand \@ifx [1]{%
 \ifx #1\expandafter \@firstoftwo
 \else \expandafter \@secondoftwo
 \fi
}%
\providecommand \natexlab [1]{#1}%
\providecommand \enquote  [1]{``#1''}%
\providecommand \bibnamefont  [1]{#1}%
\providecommand \bibfnamefont [1]{#1}%
\providecommand \citenamefont [1]{#1}%
\providecommand \href@noop [0]{\@secondoftwo}%
\providecommand \href [0]{\begingroup \@sanitize@url \@href}%
\providecommand \@href[1]{\@@startlink{#1}\@@href}%
\providecommand \@@href[1]{\endgroup#1\@@endlink}%
\providecommand \@sanitize@url [0]{\catcode `\\12\catcode `\$12\catcode `\&12\catcode `\#12\catcode `\^12\catcode `\_12\catcode `\%12\relax}%
\providecommand \@@startlink[1]{}%
\providecommand \@@endlink[0]{}%
\providecommand \url  [0]{\begingroup\@sanitize@url \@url }%
\providecommand \@url [1]{\endgroup\@href {#1}{\urlprefix }}%
\providecommand \urlprefix  [0]{URL }%
\providecommand \Eprint [0]{\href }%
\providecommand \doibase [0]{https://doi.org/}%
\providecommand \selectlanguage [0]{\@gobble}%
\providecommand \bibinfo  [0]{\@secondoftwo}%
\providecommand \bibfield  [0]{\@secondoftwo}%
\providecommand \translation [1]{[#1]}%
\providecommand \BibitemOpen [0]{}%
\providecommand \bibitemStop [0]{}%
\providecommand \bibitemNoStop [0]{.\EOS\space}%
\providecommand \EOS [0]{\spacefactor3000\relax}%
\providecommand \BibitemShut  [1]{\csname bibitem#1\endcsname}%
\let\auto@bib@innerbib\@empty
\bibitem [{\citenamefont {Feynman}(1982)}]{cite-key}%
  \BibitemOpen
  \bibfield  {author} {\bibinfo {author} {\bibfnamefont {R.~P.}\ \bibnamefont {Feynman}},\ }\bibfield  {title} {\bibinfo {title} {Simulating physics with computers},\ }\href {https://doi.org/10.1007/BF02650179} {\bibfield  {journal} {\bibinfo  {journal} {International Journal of Theoretical Physics}\ }\textbf {\bibinfo {volume} {21}},\ \bibinfo {pages} {467} (\bibinfo {year} {1982})}\BibitemShut {NoStop}%
\bibitem [{\citenamefont {Jordan}\ \emph {et~al.}(2012)\citenamefont {Jordan}, \citenamefont {Lee},\ and\ \citenamefont {Preskill}}]{doi:10.1126/science.1217069}%
  \BibitemOpen
  \bibfield  {author} {\bibinfo {author} {\bibfnamefont {S.~P.}\ \bibnamefont {Jordan}}, \bibinfo {author} {\bibfnamefont {K.~S.~M.}\ \bibnamefont {Lee}},\ and\ \bibinfo {author} {\bibfnamefont {J.}~\bibnamefont {Preskill}},\ }\bibfield  {title} {\bibinfo {title} {Quantum algorithms for quantum field theories},\ }\href {https://doi.org/10.1126/science.1217069} {\bibfield  {journal} {\bibinfo  {journal} {Science}\ }\textbf {\bibinfo {volume} {336}},\ \bibinfo {pages} {1130} (\bibinfo {year} {2012})},\ \Eprint {https://arxiv.org/abs/https://www.science.org/doi/pdf/10.1126/science.1217069} {https://www.science.org/doi/pdf/10.1126/science.1217069} \BibitemShut {NoStop}%
\bibitem [{\citenamefont {Bauer}\ \emph {et~al.}(2023)\citenamefont {Bauer}, \citenamefont {Davoudi}, \citenamefont {Balantekin}, \citenamefont {Bhattacharya}, \citenamefont {Carena}, \citenamefont {de~Jong}, \citenamefont {Draper}, \citenamefont {El-Khadra}, \citenamefont {Gemelke}, \citenamefont {Hanada}, \citenamefont {Kharzeev}, \citenamefont {Lamm}, \citenamefont {Li}, \citenamefont {Liu}, \citenamefont {Lukin}, \citenamefont {Meurice}, \citenamefont {Monroe}, \citenamefont {Nachman}, \citenamefont {Pagano}, \citenamefont {Preskill}, \citenamefont {Rinaldi}, \citenamefont {Roggero}, \citenamefont {Santiago}, \citenamefont {Savage}, \citenamefont {Siddiqi}, \citenamefont {Siopsis}, \citenamefont {Van~Zanten}, \citenamefont {Wiebe}, \citenamefont {Yamauchi}, \citenamefont {Yeter-Aydeniz},\ and\ \citenamefont {Zorzetti}}]{PRXQuantum.4.027001}%
  \BibitemOpen
  \bibfield  {author} {\bibinfo {author} {\bibfnamefont {C.~W.}\ \bibnamefont {Bauer}}, \bibinfo {author} {\bibfnamefont {Z.}~\bibnamefont {Davoudi}}, \bibinfo {author} {\bibfnamefont {A.~B.}\ \bibnamefont {Balantekin}}, \bibinfo {author} {\bibfnamefont {T.}~\bibnamefont {Bhattacharya}}, \bibinfo {author} {\bibfnamefont {M.}~\bibnamefont {Carena}}, \bibinfo {author} {\bibfnamefont {W.~A.}\ \bibnamefont {de~Jong}}, \bibinfo {author} {\bibfnamefont {P.}~\bibnamefont {Draper}}, \bibinfo {author} {\bibfnamefont {A.}~\bibnamefont {El-Khadra}}, \bibinfo {author} {\bibfnamefont {N.}~\bibnamefont {Gemelke}}, \bibinfo {author} {\bibfnamefont {M.}~\bibnamefont {Hanada}}, \bibinfo {author} {\bibfnamefont {D.}~\bibnamefont {Kharzeev}}, \bibinfo {author} {\bibfnamefont {H.}~\bibnamefont {Lamm}}, \bibinfo {author} {\bibfnamefont {Y.-Y.}\ \bibnamefont {Li}}, \bibinfo {author} {\bibfnamefont {J.}~\bibnamefont {Liu}}, \bibinfo {author} {\bibfnamefont {M.}~\bibnamefont {Lukin}}, \bibinfo {author} {\bibfnamefont
  {Y.}~\bibnamefont {Meurice}}, \bibinfo {author} {\bibfnamefont {C.}~\bibnamefont {Monroe}}, \bibinfo {author} {\bibfnamefont {B.}~\bibnamefont {Nachman}}, \bibinfo {author} {\bibfnamefont {G.}~\bibnamefont {Pagano}}, \bibinfo {author} {\bibfnamefont {J.}~\bibnamefont {Preskill}}, \bibinfo {author} {\bibfnamefont {E.}~\bibnamefont {Rinaldi}}, \bibinfo {author} {\bibfnamefont {A.}~\bibnamefont {Roggero}}, \bibinfo {author} {\bibfnamefont {D.~I.}\ \bibnamefont {Santiago}}, \bibinfo {author} {\bibfnamefont {M.~J.}\ \bibnamefont {Savage}}, \bibinfo {author} {\bibfnamefont {I.}~\bibnamefont {Siddiqi}}, \bibinfo {author} {\bibfnamefont {G.}~\bibnamefont {Siopsis}}, \bibinfo {author} {\bibfnamefont {D.}~\bibnamefont {Van~Zanten}}, \bibinfo {author} {\bibfnamefont {N.}~\bibnamefont {Wiebe}}, \bibinfo {author} {\bibfnamefont {Y.}~\bibnamefont {Yamauchi}}, \bibinfo {author} {\bibfnamefont {K.}~\bibnamefont {Yeter-Aydeniz}},\ and\ \bibinfo {author} {\bibfnamefont {S.}~\bibnamefont {Zorzetti}},\ }\bibfield  {title}
  {\bibinfo {title} {Quantum simulation for high-energy physics},\ }\href {https://doi.org/10.1103/PRXQuantum.4.027001} {\bibfield  {journal} {\bibinfo  {journal} {PRX Quantum}\ }\textbf {\bibinfo {volume} {4}},\ \bibinfo {pages} {027001} (\bibinfo {year} {2023})}\BibitemShut {NoStop}%
\bibitem [{\citenamefont {Cervia}\ \emph {et~al.}(2021)\citenamefont {Cervia}, \citenamefont {Balantekin}, \citenamefont {Coppersmith}, \citenamefont {Johnson}, \citenamefont {Love}, \citenamefont {Poole}, \citenamefont {Robbins},\ and\ \citenamefont {Saffman}}]{PhysRevC.104.024305}%
  \BibitemOpen
  \bibfield  {author} {\bibinfo {author} {\bibfnamefont {M.~J.}\ \bibnamefont {Cervia}}, \bibinfo {author} {\bibfnamefont {A.~B.}\ \bibnamefont {Balantekin}}, \bibinfo {author} {\bibfnamefont {S.~N.}\ \bibnamefont {Coppersmith}}, \bibinfo {author} {\bibfnamefont {C.~W.}\ \bibnamefont {Johnson}}, \bibinfo {author} {\bibfnamefont {P.~J.}\ \bibnamefont {Love}}, \bibinfo {author} {\bibfnamefont {C.}~\bibnamefont {Poole}}, \bibinfo {author} {\bibfnamefont {K.}~\bibnamefont {Robbins}},\ and\ \bibinfo {author} {\bibfnamefont {M.}~\bibnamefont {Saffman}},\ }\bibfield  {title} {\bibinfo {title} {Lipkin model on a quantum computer},\ }\href {https://doi.org/10.1103/PhysRevC.104.024305} {\bibfield  {journal} {\bibinfo  {journal} {Phys. Rev. C}\ }\textbf {\bibinfo {volume} {104}},\ \bibinfo {pages} {024305} (\bibinfo {year} {2021})}\BibitemShut {NoStop}%
\bibitem [{\citenamefont {Hensgens}\ \emph {et~al.}(2017)\citenamefont {Hensgens}, \citenamefont {Fujita}, \citenamefont {Janssen}, \citenamefont {Li}, \citenamefont {Van~Diepen}, \citenamefont {Reichl}, \citenamefont {Wegscheider}, \citenamefont {Das~Sarma},\ and\ \citenamefont {Vandersypen}}]{Hensgens_2017}%
  \BibitemOpen
  \bibfield  {author} {\bibinfo {author} {\bibfnamefont {T.}~\bibnamefont {Hensgens}}, \bibinfo {author} {\bibfnamefont {T.}~\bibnamefont {Fujita}}, \bibinfo {author} {\bibfnamefont {L.}~\bibnamefont {Janssen}}, \bibinfo {author} {\bibfnamefont {X.}~\bibnamefont {Li}}, \bibinfo {author} {\bibfnamefont {C.~J.}\ \bibnamefont {Van~Diepen}}, \bibinfo {author} {\bibfnamefont {C.}~\bibnamefont {Reichl}}, \bibinfo {author} {\bibfnamefont {W.}~\bibnamefont {Wegscheider}}, \bibinfo {author} {\bibfnamefont {S.}~\bibnamefont {Das~Sarma}},\ and\ \bibinfo {author} {\bibfnamefont {L.~M.~K.}\ \bibnamefont {Vandersypen}},\ }\bibfield  {title} {\bibinfo {title} {Quantum simulation of a fermi–hubbard model using a semiconductor quantum dot array},\ }\href {https://doi.org/10.1038/nature23022} {\bibfield  {journal} {\bibinfo  {journal} {Nature}\ }\textbf {\bibinfo {volume} {548}},\ \bibinfo {pages} {70–73} (\bibinfo {year} {2017})}\BibitemShut {NoStop}%
\bibitem [{\citenamefont {Cervera-Lierta}(2018)}]{Cervera_Lierta_2018}%
  \BibitemOpen
  \bibfield  {author} {\bibinfo {author} {\bibfnamefont {A.}~\bibnamefont {Cervera-Lierta}},\ }\bibfield  {title} {\bibinfo {title} {Exact ising model simulation on a quantum computer},\ }\href {https://doi.org/10.22331/q-2018-12-21-114} {\bibfield  {journal} {\bibinfo  {journal} {Quantum}\ }\textbf {\bibinfo {volume} {2}},\ \bibinfo {pages} {114} (\bibinfo {year} {2018})}\BibitemShut {NoStop}%
\bibitem [{\citenamefont {Geller}\ \emph {et~al.}(2022)\citenamefont {Geller}, \citenamefont {Arrasmith}, \citenamefont {Holmes}, \citenamefont {Yan}, \citenamefont {Coles},\ and\ \citenamefont {Sornborger}}]{Geller_2022}%
  \BibitemOpen
  \bibfield  {author} {\bibinfo {author} {\bibfnamefont {M.~R.}\ \bibnamefont {Geller}}, \bibinfo {author} {\bibfnamefont {A.}~\bibnamefont {Arrasmith}}, \bibinfo {author} {\bibfnamefont {Z.}~\bibnamefont {Holmes}}, \bibinfo {author} {\bibfnamefont {B.}~\bibnamefont {Yan}}, \bibinfo {author} {\bibfnamefont {P.~J.}\ \bibnamefont {Coles}},\ and\ \bibinfo {author} {\bibfnamefont {A.}~\bibnamefont {Sornborger}},\ }\bibfield  {title} {\bibinfo {title} {Quantum simulation of operator spreading in the chaotic ising model},\ }\bibfield  {journal} {\bibinfo  {journal} {Physical Review E}\ }\textbf {\bibinfo {volume} {105}},\ \href {https://doi.org/10.1103/physreve.105.035302} {10.1103/physreve.105.035302} (\bibinfo {year} {2022})\BibitemShut {NoStop}%
\bibitem [{\citenamefont {Macridin}\ \emph {et~al.}(2018)\citenamefont {Macridin}, \citenamefont {Spentzouris}, \citenamefont {Amundson},\ and\ \citenamefont {Harnik}}]{Macridin_2018}%
  \BibitemOpen
  \bibfield  {author} {\bibinfo {author} {\bibfnamefont {A.}~\bibnamefont {Macridin}}, \bibinfo {author} {\bibfnamefont {P.}~\bibnamefont {Spentzouris}}, \bibinfo {author} {\bibfnamefont {J.}~\bibnamefont {Amundson}},\ and\ \bibinfo {author} {\bibfnamefont {R.}~\bibnamefont {Harnik}},\ }\bibfield  {title} {\bibinfo {title} {Electron-phonon systems on a universal quantum computer},\ }\bibfield  {journal} {\bibinfo  {journal} {Physical Review Letters}\ }\textbf {\bibinfo {volume} {121}},\ \href {https://doi.org/10.1103/physrevlett.121.110504} {10.1103/physrevlett.121.110504} (\bibinfo {year} {2018})\BibitemShut {NoStop}%
\bibitem [{\citenamefont {Li}\ \emph {et~al.}(2020)\citenamefont {Li}, \citenamefont {Venitucci},\ and\ \citenamefont {Niquet}}]{Li_2020}%
  \BibitemOpen
  \bibfield  {author} {\bibinfo {author} {\bibfnamefont {J.}~\bibnamefont {Li}}, \bibinfo {author} {\bibfnamefont {B.}~\bibnamefont {Venitucci}},\ and\ \bibinfo {author} {\bibfnamefont {Y.-M.}\ \bibnamefont {Niquet}},\ }\bibfield  {title} {\bibinfo {title} {Hole-phonon interactions in quantum dots: Effects of phonon confinement and encapsulation materials on spin-orbit qubits},\ }\bibfield  {journal} {\bibinfo  {journal} {Physical Review B}\ }\textbf {\bibinfo {volume} {102}},\ \href {https://doi.org/10.1103/physrevb.102.075415} {10.1103/physrevb.102.075415} (\bibinfo {year} {2020})\BibitemShut {NoStop}%
\bibitem [{\citenamefont {Cai}\ \emph {et~al.}(2023)\citenamefont {Cai}, \citenamefont {Babbush}, \citenamefont {Benjamin}, \citenamefont {Endo}, \citenamefont {Huggins}, \citenamefont {Li}, \citenamefont {McClean},\ and\ \citenamefont {O'Brien}}]{RevModPhys.95.045005}%
  \BibitemOpen
  \bibfield  {author} {\bibinfo {author} {\bibfnamefont {Z.}~\bibnamefont {Cai}}, \bibinfo {author} {\bibfnamefont {R.}~\bibnamefont {Babbush}}, \bibinfo {author} {\bibfnamefont {S.~C.}\ \bibnamefont {Benjamin}}, \bibinfo {author} {\bibfnamefont {S.}~\bibnamefont {Endo}}, \bibinfo {author} {\bibfnamefont {W.~J.}\ \bibnamefont {Huggins}}, \bibinfo {author} {\bibfnamefont {Y.}~\bibnamefont {Li}}, \bibinfo {author} {\bibfnamefont {J.~R.}\ \bibnamefont {McClean}},\ and\ \bibinfo {author} {\bibfnamefont {T.~E.}\ \bibnamefont {O'Brien}},\ }\bibfield  {title} {\bibinfo {title} {Quantum error mitigation},\ }\href {https://doi.org/10.1103/RevModPhys.95.045005} {\bibfield  {journal} {\bibinfo  {journal} {Rev. Mod. Phys.}\ }\textbf {\bibinfo {volume} {95}},\ \bibinfo {pages} {045005} (\bibinfo {year} {2023})}\BibitemShut {NoStop}%
\bibitem [{\citenamefont {van~den Berg}\ \emph {et~al.}(2023)\citenamefont {van~den Berg}, \citenamefont {Minev}, \citenamefont {Kandala},\ and\ \citenamefont {Temme}}]{PEC}%
  \BibitemOpen
  \bibfield  {author} {\bibinfo {author} {\bibfnamefont {E.}~\bibnamefont {van~den Berg}}, \bibinfo {author} {\bibfnamefont {Z.~K.}\ \bibnamefont {Minev}}, \bibinfo {author} {\bibfnamefont {A.}~\bibnamefont {Kandala}},\ and\ \bibinfo {author} {\bibfnamefont {K.}~\bibnamefont {Temme}},\ }\bibfield  {title} {\bibinfo {title} {Probabilistic error cancellation with sparse pauli–lindblad models on noisy quantum processors},\ }\href {https://doi.org/10.1038/s41567-023-02042-2} {\bibfield  {journal} {\bibinfo  {journal} {Nature Physics}\ }\textbf {\bibinfo {volume} {19}},\ \bibinfo {pages} {1116–1121} (\bibinfo {year} {2023})}\BibitemShut {NoStop}%
\bibitem [{\citenamefont {Takagi}\ \emph {et~al.}(2022)\citenamefont {Takagi}, \citenamefont {Endo}, \citenamefont {Minagawa} \emph {et~al.}}]{Takagi2022}%
  \BibitemOpen
  \bibfield  {author} {\bibinfo {author} {\bibfnamefont {R.}~\bibnamefont {Takagi}}, \bibinfo {author} {\bibfnamefont {S.}~\bibnamefont {Endo}}, \bibinfo {author} {\bibfnamefont {S.}~\bibnamefont {Minagawa}}, \emph {et~al.},\ }\bibfield  {title} {\bibinfo {title} {Fundamental limits of quantum error mitigation},\ }\bibfield  {journal} {\bibinfo  {journal} {npj Quantum Information}\ }\textbf {\bibinfo {volume} {8}},\ \href {https://doi.org/10.1038/s41534-022-00618-z} {10.1038/s41534-022-00618-z} (\bibinfo {year} {2022})\BibitemShut {NoStop}%
\bibitem [{\citenamefont {Giurgica-Tiron}\ \emph {et~al.}(2020)\citenamefont {Giurgica-Tiron}, \citenamefont {Hindy}, \citenamefont {LaRose}, \citenamefont {Mari},\ and\ \citenamefont {Zeng}}]{Giurgica_Tiron_2020}%
  \BibitemOpen
  \bibfield  {author} {\bibinfo {author} {\bibfnamefont {T.}~\bibnamefont {Giurgica-Tiron}}, \bibinfo {author} {\bibfnamefont {Y.}~\bibnamefont {Hindy}}, \bibinfo {author} {\bibfnamefont {R.}~\bibnamefont {LaRose}}, \bibinfo {author} {\bibfnamefont {A.}~\bibnamefont {Mari}},\ and\ \bibinfo {author} {\bibfnamefont {W.~J.}\ \bibnamefont {Zeng}},\ }\bibfield  {title} {\bibinfo {title} {Digital zero noise extrapolation for quantum error mitigation},\ }in\ \href {https://doi.org/10.1109/qce49297.2020.00045} {\emph {\bibinfo {booktitle} {2020 IEEE International Conference on Quantum Computing and Engineering (QCE)}}}\ (\bibinfo  {publisher} {IEEE},\ \bibinfo {year} {2020})\BibitemShut {NoStop}%
\bibitem [{\citenamefont {Feng}\ and\ \citenamefont {Ruan}(2016{\natexlab{a}})}]{Feng_2016}%
  \BibitemOpen
  \bibfield  {author} {\bibinfo {author} {\bibfnamefont {T.}~\bibnamefont {Feng}}\ and\ \bibinfo {author} {\bibfnamefont {X.}~\bibnamefont {Ruan}},\ }\bibfield  {title} {\bibinfo {title} {Quantum mechanical prediction of four-phonon scattering rates and reduced thermal conductivity of solids},\ }\bibfield  {journal} {\bibinfo  {journal} {Physical Review B}\ }\textbf {\bibinfo {volume} {93}},\ \href {https://doi.org/10.1103/physrevb.93.045202} {10.1103/physrevb.93.045202} (\bibinfo {year} {2016}{\natexlab{a}})\BibitemShut {NoStop}%
\bibitem [{\citenamefont {Ghosh}\ \emph {et~al.}(2023)\citenamefont {Ghosh}, \citenamefont {Kusiak},\ and\ \citenamefont {Battaglia}}]{Ghosh_2023}%
  \BibitemOpen
  \bibfield  {author} {\bibinfo {author} {\bibfnamefont {K.}~\bibnamefont {Ghosh}}, \bibinfo {author} {\bibfnamefont {A.}~\bibnamefont {Kusiak}},\ and\ \bibinfo {author} {\bibfnamefont {J.-L.}\ \bibnamefont {Battaglia}},\ }\bibfield  {title} {\bibinfo {title} {Significant four-phonon scattering and its heat transfer implications in crystalline},\ }\bibfield  {journal} {\bibinfo  {journal} {Physical Review B}\ }\textbf {\bibinfo {volume} {108}},\ \href {https://doi.org/10.1103/physrevb.108.214309} {10.1103/physrevb.108.214309} (\bibinfo {year} {2023})\BibitemShut {NoStop}%
\bibitem [{\citenamefont {Li}\ \emph {et~al.}(2014)\citenamefont {Li}, \citenamefont {Carrete}, \citenamefont {Katcho},\ and\ \citenamefont {Mingo}}]{ShengBTE_2014}%
  \BibitemOpen
  \bibfield  {author} {\bibinfo {author} {\bibfnamefont {W.}~\bibnamefont {Li}}, \bibinfo {author} {\bibfnamefont {J.}~\bibnamefont {Carrete}}, \bibinfo {author} {\bibfnamefont {N.~A.}\ \bibnamefont {Katcho}},\ and\ \bibinfo {author} {\bibfnamefont {N.}~\bibnamefont {Mingo}},\ }\bibfield  {title} {\bibinfo {title} {{ShengBTE:} a solver of the {B}oltzmann transport equation for phonons},\ }\href {https://doi.org/10.1016/j.cpc.2014.02.015} {\bibfield  {journal} {\bibinfo  {journal} {Comp. Phys. Commun.}\ }\textbf {\bibinfo {volume} {185}},\ \bibinfo {pages} {1747–1758} (\bibinfo {year} {2014})}\BibitemShut {NoStop}%
\bibitem [{\citenamefont {Li}\ \emph {et~al.}(2012)\citenamefont {Li}, \citenamefont {Lindsay}, \citenamefont {Broido}, \citenamefont {Stewart},\ and\ \citenamefont {Mingo}}]{PhysRevB.86.174307}%
  \BibitemOpen
  \bibfield  {author} {\bibinfo {author} {\bibfnamefont {W.}~\bibnamefont {Li}}, \bibinfo {author} {\bibfnamefont {L.}~\bibnamefont {Lindsay}}, \bibinfo {author} {\bibfnamefont {D.~A.}\ \bibnamefont {Broido}}, \bibinfo {author} {\bibfnamefont {D.~A.}\ \bibnamefont {Stewart}},\ and\ \bibinfo {author} {\bibfnamefont {N.}~\bibnamefont {Mingo}},\ }\bibfield  {title} {\bibinfo {title} {Thermal conductivity of bulk and nanowire mg${}_{2}$si${}_{x}$sn${}_{1\ensuremath{-}x}$ alloys from first principles},\ }\href {https://doi.org/10.1103/PhysRevB.86.174307} {\bibfield  {journal} {\bibinfo  {journal} {Phys. Rev. B}\ }\textbf {\bibinfo {volume} {86}},\ \bibinfo {pages} {174307} (\bibinfo {year} {2012})}\BibitemShut {NoStop}%
\bibitem [{\citenamefont {Tilly}\ \emph {et~al.}(2022)\citenamefont {Tilly}, \citenamefont {Chen}, \citenamefont {Cao}, \citenamefont {Picozzi}, \citenamefont {Setia}, \citenamefont {Li}, \citenamefont {Grant}, \citenamefont {Wossnig}, \citenamefont {Rungger}, \citenamefont {Booth},\ and\ \citenamefont {Tennyson}}]{Tilly_2022}%
  \BibitemOpen
  \bibfield  {author} {\bibinfo {author} {\bibfnamefont {J.}~\bibnamefont {Tilly}}, \bibinfo {author} {\bibfnamefont {H.}~\bibnamefont {Chen}}, \bibinfo {author} {\bibfnamefont {S.}~\bibnamefont {Cao}}, \bibinfo {author} {\bibfnamefont {D.}~\bibnamefont {Picozzi}}, \bibinfo {author} {\bibfnamefont {K.}~\bibnamefont {Setia}}, \bibinfo {author} {\bibfnamefont {Y.}~\bibnamefont {Li}}, \bibinfo {author} {\bibfnamefont {E.}~\bibnamefont {Grant}}, \bibinfo {author} {\bibfnamefont {L.}~\bibnamefont {Wossnig}}, \bibinfo {author} {\bibfnamefont {I.}~\bibnamefont {Rungger}}, \bibinfo {author} {\bibfnamefont {G.~H.}\ \bibnamefont {Booth}},\ and\ \bibinfo {author} {\bibfnamefont {J.}~\bibnamefont {Tennyson}},\ }\bibfield  {title} {\bibinfo {title} {The variational quantum eigensolver: A review of methods and best practices},\ }\href {https://doi.org/10.1016/j.physrep.2022.08.003} {\bibfield  {journal} {\bibinfo  {journal} {Physics Reports}\ }\textbf {\bibinfo {volume} {986}},\ \bibinfo {pages} {1–128} (\bibinfo {year}
  {2022})}\BibitemShut {NoStop}%
\bibitem [{\citenamefont {Daley}(2023)}]{Daley2023}%
  \BibitemOpen
  \bibfield  {author} {\bibinfo {author} {\bibfnamefont {A.}~\bibnamefont {Daley}},\ }\bibfield  {title} {\bibinfo {title} {Twenty-five years of analogue quantum simulation},\ }\href {https://doi.org/10.1038/s42254-023-00666-0} {\bibfield  {journal} {\bibinfo  {journal} {Nature Reviews Physics}\ }\textbf {\bibinfo {volume} {5}},\ \bibinfo {pages} {702} (\bibinfo {year} {2023})}\BibitemShut {NoStop}%
\bibitem [{\citenamefont {Zhang}\ \emph {et~al.}(2023)\citenamefont {Zhang}, \citenamefont {Kim}, \citenamefont {Mark}, \citenamefont {Choi},\ and\ \citenamefont {Painter}}]{doi:10.1126/science.ade7651}%
  \BibitemOpen
  \bibfield  {author} {\bibinfo {author} {\bibfnamefont {X.}~\bibnamefont {Zhang}}, \bibinfo {author} {\bibfnamefont {E.}~\bibnamefont {Kim}}, \bibinfo {author} {\bibfnamefont {D.~K.}\ \bibnamefont {Mark}}, \bibinfo {author} {\bibfnamefont {S.}~\bibnamefont {Choi}},\ and\ \bibinfo {author} {\bibfnamefont {O.}~\bibnamefont {Painter}},\ }\bibfield  {title} {\bibinfo {title} {A superconducting quantum simulator based on a photonic-bandgap metamaterial},\ }\href {https://doi.org/10.1126/science.ade7651} {\bibfield  {journal} {\bibinfo  {journal} {Science}\ }\textbf {\bibinfo {volume} {379}},\ \bibinfo {pages} {278} (\bibinfo {year} {2023})}\BibitemShut {NoStop}%
\bibitem [{\citenamefont {Weckesser}\ \emph {et~al.}(2024)\citenamefont {Weckesser}, \citenamefont {Srakaew}, \citenamefont {Blatz}, \citenamefont {Wei}, \citenamefont {Adler}, \citenamefont {Agrawal}, \citenamefont {Bohrdt}, \citenamefont {Bloch},\ and\ \citenamefont {Zeiher}}]{weckesser2024realizationrydbergdressedextendedbose}%
  \BibitemOpen
  \bibfield  {author} {\bibinfo {author} {\bibfnamefont {P.}~\bibnamefont {Weckesser}}, \bibinfo {author} {\bibfnamefont {K.}~\bibnamefont {Srakaew}}, \bibinfo {author} {\bibfnamefont {T.}~\bibnamefont {Blatz}}, \bibinfo {author} {\bibfnamefont {D.}~\bibnamefont {Wei}}, \bibinfo {author} {\bibfnamefont {D.}~\bibnamefont {Adler}}, \bibinfo {author} {\bibfnamefont {S.}~\bibnamefont {Agrawal}}, \bibinfo {author} {\bibfnamefont {A.}~\bibnamefont {Bohrdt}}, \bibinfo {author} {\bibfnamefont {I.}~\bibnamefont {Bloch}},\ and\ \bibinfo {author} {\bibfnamefont {J.}~\bibnamefont {Zeiher}},\ }\href {https://arxiv.org/abs/2405.20128} {\bibinfo {title} {Realization of a rydberg-dressed extended bose hubbard model}} (\bibinfo {year} {2024}),\ \Eprint {https://arxiv.org/abs/2405.20128} {arXiv:2405.20128 [cond-mat.quant-gas]} \BibitemShut {NoStop}%
\bibitem [{\citenamefont {Anand}\ \emph {et~al.}(2022)\citenamefont {Anand}, \citenamefont {Schleich}, \citenamefont {Alperin-Lea}, \citenamefont {Jensen}, \citenamefont {Sim}, \citenamefont {Díaz-Tinoco}, \citenamefont {Kottmann}, \citenamefont {Degroote}, \citenamefont {Izmaylov},\ and\ \citenamefont {Aspuru-Guzik}}]{Anand_2022}%
  \BibitemOpen
  \bibfield  {author} {\bibinfo {author} {\bibfnamefont {A.}~\bibnamefont {Anand}}, \bibinfo {author} {\bibfnamefont {P.}~\bibnamefont {Schleich}}, \bibinfo {author} {\bibfnamefont {S.}~\bibnamefont {Alperin-Lea}}, \bibinfo {author} {\bibfnamefont {P.~W.~K.}\ \bibnamefont {Jensen}}, \bibinfo {author} {\bibfnamefont {S.}~\bibnamefont {Sim}}, \bibinfo {author} {\bibfnamefont {M.}~\bibnamefont {Díaz-Tinoco}}, \bibinfo {author} {\bibfnamefont {J.~S.}\ \bibnamefont {Kottmann}}, \bibinfo {author} {\bibfnamefont {M.}~\bibnamefont {Degroote}}, \bibinfo {author} {\bibfnamefont {A.~F.}\ \bibnamefont {Izmaylov}},\ and\ \bibinfo {author} {\bibfnamefont {A.}~\bibnamefont {Aspuru-Guzik}},\ }\bibfield  {title} {\bibinfo {title} {A quantum computing view on unitary coupled cluster theory},\ }\href {https://doi.org/10.1039/d1cs00932j} {\bibfield  {journal} {\bibinfo  {journal} {Chemical Society Reviews}\ }\textbf {\bibinfo {volume} {51}},\ \bibinfo {pages} {1659–1684} (\bibinfo {year} {2022})}\BibitemShut {NoStop}%
\bibitem [{\citenamefont {Romero}\ \emph {et~al.}(2018)\citenamefont {Romero}, \citenamefont {Babbush}, \citenamefont {McClean}, \citenamefont {Hempel}, \citenamefont {Love},\ and\ \citenamefont {Aspuru-Guzik}}]{romero2018strategies}%
  \BibitemOpen
  \bibfield  {author} {\bibinfo {author} {\bibfnamefont {J.}~\bibnamefont {Romero}}, \bibinfo {author} {\bibfnamefont {R.}~\bibnamefont {Babbush}}, \bibinfo {author} {\bibfnamefont {J.~R.}\ \bibnamefont {McClean}}, \bibinfo {author} {\bibfnamefont {C.}~\bibnamefont {Hempel}}, \bibinfo {author} {\bibfnamefont {P.}~\bibnamefont {Love}},\ and\ \bibinfo {author} {\bibfnamefont {A.}~\bibnamefont {Aspuru-Guzik}},\ }\href@noop {} {\bibinfo {title} {Strategies for quantum computing molecular energies using the unitary coupled cluster ansatz}} (\bibinfo {year} {2018}),\ \Eprint {https://arxiv.org/abs/1701.02691} {arXiv:1701.02691 [quant-ph]} \BibitemShut {NoStop}%
\bibitem [{\citenamefont {Fedorov}\ \emph {et~al.}(2022)\citenamefont {Fedorov}, \citenamefont {Peng}, \citenamefont {Govind},\ and\ \citenamefont {Alexeev}}]{VQE_method}%
  \BibitemOpen
  \bibfield  {author} {\bibinfo {author} {\bibfnamefont {D.~A.}\ \bibnamefont {Fedorov}}, \bibinfo {author} {\bibfnamefont {B.}~\bibnamefont {Peng}}, \bibinfo {author} {\bibfnamefont {N.}~\bibnamefont {Govind}},\ and\ \bibinfo {author} {\bibfnamefont {Y.}~\bibnamefont {Alexeev}},\ }\bibfield  {title} {\bibinfo {title} {Vqe method: a short survey and recent developments},\ }\href {https://doi.org/10.1186/s41313-021-00032-6} {\bibfield  {journal} {\bibinfo  {journal} {Materials Theory}\ }\textbf {\bibinfo {volume} {6}},\ \bibinfo {pages} {2} (\bibinfo {year} {2022})}\BibitemShut {NoStop}%
\bibitem [{\citenamefont {Katz}\ and\ \citenamefont {Monroe}(2023)}]{PhysRevLett.131.033604}%
  \BibitemOpen
  \bibfield  {author} {\bibinfo {author} {\bibfnamefont {O.}~\bibnamefont {Katz}}\ and\ \bibinfo {author} {\bibfnamefont {C.}~\bibnamefont {Monroe}},\ }\bibfield  {title} {\bibinfo {title} {Programmable quantum simulations of bosonic systems with trapped ions},\ }\href {https://doi.org/10.1103/PhysRevLett.131.033604} {\bibfield  {journal} {\bibinfo  {journal} {Phys. Rev. Lett.}\ }\textbf {\bibinfo {volume} {131}},\ \bibinfo {pages} {033604} (\bibinfo {year} {2023})}\BibitemShut {NoStop}%
\bibitem [{\citenamefont {Busnaina}\ \emph {et~al.}(2024)\citenamefont {Busnaina}, \citenamefont {Shi}, \citenamefont {McDonald} \emph {et~al.}}]{busnaina2024quantum}%
  \BibitemOpen
  \bibfield  {author} {\bibinfo {author} {\bibfnamefont {J.}~\bibnamefont {Busnaina}}, \bibinfo {author} {\bibfnamefont {Z.}~\bibnamefont {Shi}}, \bibinfo {author} {\bibfnamefont {A.}~\bibnamefont {McDonald}}, \emph {et~al.},\ }\bibfield  {title} {\bibinfo {title} {Quantum simulation of the bosonic kitaev chain},\ }\href {https://doi.org/10.1038/s41467-024-47186-8} {\bibfield  {journal} {\bibinfo  {journal} {Nature Communications}\ }\textbf {\bibinfo {volume} {15}},\ \bibinfo {pages} {3065} (\bibinfo {year} {2024})}\BibitemShut {NoStop}%
\bibitem [{\citenamefont {Javadi-Abhari}\ \emph {et~al.}(2024)\citenamefont {Javadi-Abhari}, \citenamefont {Treinish}, \citenamefont {Krsulich}, \citenamefont {Wood}, \citenamefont {Lishman}, \citenamefont {Gacon}, \citenamefont {Martiel}, \citenamefont {Nation}, \citenamefont {Bishop}, \citenamefont {Cross}, \citenamefont {Johnson},\ and\ \citenamefont {Gambetta}}]{qiskit2024}%
  \BibitemOpen
  \bibfield  {author} {\bibinfo {author} {\bibfnamefont {A.}~\bibnamefont {Javadi-Abhari}}, \bibinfo {author} {\bibfnamefont {M.}~\bibnamefont {Treinish}}, \bibinfo {author} {\bibfnamefont {K.}~\bibnamefont {Krsulich}}, \bibinfo {author} {\bibfnamefont {C.~J.}\ \bibnamefont {Wood}}, \bibinfo {author} {\bibfnamefont {J.}~\bibnamefont {Lishman}}, \bibinfo {author} {\bibfnamefont {J.}~\bibnamefont {Gacon}}, \bibinfo {author} {\bibfnamefont {S.}~\bibnamefont {Martiel}}, \bibinfo {author} {\bibfnamefont {P.~D.}\ \bibnamefont {Nation}}, \bibinfo {author} {\bibfnamefont {L.~S.}\ \bibnamefont {Bishop}}, \bibinfo {author} {\bibfnamefont {A.~W.}\ \bibnamefont {Cross}}, \bibinfo {author} {\bibfnamefont {B.~R.}\ \bibnamefont {Johnson}},\ and\ \bibinfo {author} {\bibfnamefont {J.~M.}\ \bibnamefont {Gambetta}},\ }\href {https://doi.org/10.48550/arXiv.2405.08810} {\bibinfo {title} {Quantum computing with {Q}iskit}} (\bibinfo {year} {2024}),\ \Eprint {https://arxiv.org/abs/2405.08810} {arXiv:2405.08810 [quant-ph]}
  \BibitemShut {NoStop}%
\bibitem [{\citenamefont {Feng}\ and\ \citenamefont {Ruan}(2016{\natexlab{b}})}]{PhysRevB.93.045202}%
  \BibitemOpen
  \bibfield  {author} {\bibinfo {author} {\bibfnamefont {T.}~\bibnamefont {Feng}}\ and\ \bibinfo {author} {\bibfnamefont {X.}~\bibnamefont {Ruan}},\ }\bibfield  {title} {\bibinfo {title} {Quantum mechanical prediction of four-phonon scattering rates and reduced thermal conductivity of solids},\ }\href {https://doi.org/10.1103/PhysRevB.93.045202} {\bibfield  {journal} {\bibinfo  {journal} {Phys. Rev. B}\ }\textbf {\bibinfo {volume} {93}},\ \bibinfo {pages} {045202} (\bibinfo {year} {2016}{\natexlab{b}})}\BibitemShut {NoStop}%
\bibitem [{\citenamefont {Han}\ \emph {et~al.}(2022)\citenamefont {Han}, \citenamefont {Yang}, \citenamefont {Li}, \citenamefont {Feng},\ and\ \citenamefont {Ruan}}]{HAN2022108179}%
  \BibitemOpen
  \bibfield  {author} {\bibinfo {author} {\bibfnamefont {Z.}~\bibnamefont {Han}}, \bibinfo {author} {\bibfnamefont {X.}~\bibnamefont {Yang}}, \bibinfo {author} {\bibfnamefont {W.}~\bibnamefont {Li}}, \bibinfo {author} {\bibfnamefont {T.}~\bibnamefont {Feng}},\ and\ \bibinfo {author} {\bibfnamefont {X.}~\bibnamefont {Ruan}},\ }\bibfield  {title} {\bibinfo {title} {Fourphonon: An extension module to shengbte for computing four-phonon scattering rates and thermal conductivity},\ }\href {https://doi.org/https://doi.org/10.1016/j.cpc.2021.108179} {\bibfield  {journal} {\bibinfo  {journal} {Computer Physics Communications}\ }\textbf {\bibinfo {volume} {270}},\ \bibinfo {pages} {108179} (\bibinfo {year} {2022})}\BibitemShut {NoStop}%
\bibitem [{\citenamefont {Gu}\ \emph {et~al.}(2020)\citenamefont {Gu}, \citenamefont {Li},\ and\ \citenamefont {Bao}}]{GU2020120165}%
  \BibitemOpen
  \bibfield  {author} {\bibinfo {author} {\bibfnamefont {X.}~\bibnamefont {Gu}}, \bibinfo {author} {\bibfnamefont {S.}~\bibnamefont {Li}},\ and\ \bibinfo {author} {\bibfnamefont {H.}~\bibnamefont {Bao}},\ }\bibfield  {title} {\bibinfo {title} {Thermal conductivity of silicon at elevated temperature: Role of four-phonon scattering and electronic heat conduction},\ }\href {https://doi.org/https://doi.org/10.1016/j.ijheatmasstransfer.2020.120165} {\bibfield  {journal} {\bibinfo  {journal} {International Journal of Heat and Mass Transfer}\ }\textbf {\bibinfo {volume} {160}},\ \bibinfo {pages} {120165} (\bibinfo {year} {2020})}\BibitemShut {NoStop}%
\bibitem [{\citenamefont {Maznev}\ and\ \citenamefont {Wright}(2014)}]{maznev2014demystifying}%
  \BibitemOpen
  \bibfield  {author} {\bibinfo {author} {\bibfnamefont {A.}~\bibnamefont {Maznev}}\ and\ \bibinfo {author} {\bibfnamefont {O.}~\bibnamefont {Wright}},\ }\bibfield  {title} {\bibinfo {title} {Demystifying umklapp vs normal scattering in lattice thermal conductivity},\ }\href@noop {} {\bibfield  {journal} {\bibinfo  {journal} {American journal of physics}\ }\textbf {\bibinfo {volume} {82}},\ \bibinfo {pages} {1062} (\bibinfo {year} {2014})}\BibitemShut {NoStop}%
\bibitem [{\citenamefont {Somma}\ \emph {et~al.}(2003)\citenamefont {Somma}, \citenamefont {Ortiz}, \citenamefont {Knill},\ and\ \citenamefont {Gubernatis}}]{Somma_2003}%
  \BibitemOpen
  \bibfield  {author} {\bibinfo {author} {\bibfnamefont {R.~D.}\ \bibnamefont {Somma}}, \bibinfo {author} {\bibfnamefont {G.}~\bibnamefont {Ortiz}}, \bibinfo {author} {\bibfnamefont {E.~H.}\ \bibnamefont {Knill}},\ and\ \bibinfo {author} {\bibfnamefont {J.}~\bibnamefont {Gubernatis}},\ }\bibfield  {title} {\bibinfo {title} {Quantum simulations of physics problems},\ }in\ \href {https://doi.org/10.1117/12.487249} {\emph {\bibinfo {booktitle} {Quantum Information and Computation}}},\ \bibinfo {editor} {edited by\ \bibinfo {editor} {\bibfnamefont {E.}~\bibnamefont {Donkor}}, \bibinfo {editor} {\bibfnamefont {A.~R.}\ \bibnamefont {Pirich}},\ and\ \bibinfo {editor} {\bibfnamefont {H.~E.}\ \bibnamefont {Brandt}}}\ (\bibinfo  {publisher} {SPIE},\ \bibinfo {year} {2003})\BibitemShut {NoStop}%
\bibitem [{\citenamefont {Sawaya}\ \emph {et~al.}(2020)\citenamefont {Sawaya}, \citenamefont {Menke}, \citenamefont {Kyaw}, \citenamefont {Johri}, \citenamefont {Aspuru-Guzik},\ and\ \citenamefont {Guerreschi}}]{Nature_bosonic}%
  \BibitemOpen
  \bibfield  {author} {\bibinfo {author} {\bibfnamefont {N.~P.~D.}\ \bibnamefont {Sawaya}}, \bibinfo {author} {\bibfnamefont {T.}~\bibnamefont {Menke}}, \bibinfo {author} {\bibfnamefont {T.~H.}\ \bibnamefont {Kyaw}}, \bibinfo {author} {\bibfnamefont {S.}~\bibnamefont {Johri}}, \bibinfo {author} {\bibfnamefont {A.}~\bibnamefont {Aspuru-Guzik}},\ and\ \bibinfo {author} {\bibfnamefont {G.~G.}\ \bibnamefont {Guerreschi}},\ }\bibfield  {title} {\bibinfo {title} {Resource-efficient digital quantum simulation of d-level systems for photonic, vibrational, and spin-s hamiltonians},\ }\href {https://doi.org/10.1038/s41534-020-0278-0} {\bibfield  {journal} {\bibinfo  {journal} {npj Quantum Information}\ }\textbf {\bibinfo {volume} {6}},\ \bibinfo {pages} {49} (\bibinfo {year} {2020})}\BibitemShut {NoStop}%
\bibitem [{\citenamefont {developers}\ and\ \citenamefont {contributors}(2023)}]{the_qiskit_nature_developers_and_contrib_2023_7828768}%
  \BibitemOpen
  \bibfield  {author} {\bibinfo {author} {\bibfnamefont {T.~Q.~N.}\ \bibnamefont {developers}}\ and\ \bibinfo {author} {\bibnamefont {contributors}},\ }\href {https://doi.org/10.5281/zenodo.7828768} {\bibinfo {title} {Qiskit nature 0.6.0}} (\bibinfo {year} {2023})\BibitemShut {NoStop}%
\bibitem [{\citenamefont {Stavenger}\ \emph {et~al.}(2022)\citenamefont {Stavenger}, \citenamefont {Crane}, \citenamefont {Smith}, \citenamefont {Kang}, \citenamefont {Girvin},\ and\ \citenamefont {Wiebe}}]{stavenger2022bosonicqiskit}%
  \BibitemOpen
  \bibfield  {author} {\bibinfo {author} {\bibfnamefont {T.~J.}\ \bibnamefont {Stavenger}}, \bibinfo {author} {\bibfnamefont {E.}~\bibnamefont {Crane}}, \bibinfo {author} {\bibfnamefont {K.}~\bibnamefont {Smith}}, \bibinfo {author} {\bibfnamefont {C.~T.}\ \bibnamefont {Kang}}, \bibinfo {author} {\bibfnamefont {S.~M.}\ \bibnamefont {Girvin}},\ and\ \bibinfo {author} {\bibfnamefont {N.}~\bibnamefont {Wiebe}},\ }\href {https://arxiv.org/abs/2209.11153} {\bibinfo {title} {Bosonic qiskit}} (\bibinfo {year} {2022}),\ \Eprint {https://arxiv.org/abs/2209.11153} {arXiv:2209.11153 [quant-ph]} \BibitemShut {NoStop}%
\bibitem [{\citenamefont {Kandala}\ \emph {et~al.}(2017)\citenamefont {Kandala}, \citenamefont {Mezzacapo}, \citenamefont {Temme}, \citenamefont {Takita}, \citenamefont {Brink}, \citenamefont {Chow},\ and\ \citenamefont {Gambetta}}]{HEA}%
  \BibitemOpen
  \bibfield  {author} {\bibinfo {author} {\bibfnamefont {A.}~\bibnamefont {Kandala}}, \bibinfo {author} {\bibfnamefont {A.}~\bibnamefont {Mezzacapo}}, \bibinfo {author} {\bibfnamefont {K.}~\bibnamefont {Temme}}, \bibinfo {author} {\bibfnamefont {M.}~\bibnamefont {Takita}}, \bibinfo {author} {\bibfnamefont {M.}~\bibnamefont {Brink}}, \bibinfo {author} {\bibfnamefont {J.~M.}\ \bibnamefont {Chow}},\ and\ \bibinfo {author} {\bibfnamefont {J.~M.}\ \bibnamefont {Gambetta}},\ }\bibfield  {title} {\bibinfo {title} {Hardware-efficient variational quantum eigensolver for small molecules and quantum magnets},\ }\href {https://doi.org/10.1038/nature23879} {\bibfield  {journal} {\bibinfo  {journal} {Nature}\ }\textbf {\bibinfo {volume} {549}},\ \bibinfo {pages} {242} (\bibinfo {year} {2017})}\BibitemShut {NoStop}%
\bibitem [{\citenamefont {Funcke}\ \emph {et~al.}(2021)\citenamefont {Funcke}, \citenamefont {Hartung}, \citenamefont {Jansen}, \citenamefont {K{\"{u}}hn},\ and\ \citenamefont {Stornati}}]{Funcke2021dimensional}%
  \BibitemOpen
  \bibfield  {author} {\bibinfo {author} {\bibfnamefont {L.}~\bibnamefont {Funcke}}, \bibinfo {author} {\bibfnamefont {T.}~\bibnamefont {Hartung}}, \bibinfo {author} {\bibfnamefont {K.}~\bibnamefont {Jansen}}, \bibinfo {author} {\bibfnamefont {S.}~\bibnamefont {K{\"{u}}hn}},\ and\ \bibinfo {author} {\bibfnamefont {P.}~\bibnamefont {Stornati}},\ }\bibfield  {title} {\bibinfo {title} {Dimensional {E}xpressivity {A}nalysis of {P}arametric {Q}uantum {C}ircuits},\ }\href {https://doi.org/10.22331/q-2021-03-29-422} {\bibfield  {journal} {\bibinfo  {journal} {{Quantum}}\ }\textbf {\bibinfo {volume} {5}},\ \bibinfo {pages} {422} (\bibinfo {year} {2021})}\BibitemShut {NoStop}%
\bibitem [{\citenamefont {Feng}\ \emph {et~al.}(2017)\citenamefont {Feng}, \citenamefont {Lindsay},\ and\ \citenamefont {Ruan}}]{PhysRevB.96.161201}%
  \BibitemOpen
  \bibfield  {author} {\bibinfo {author} {\bibfnamefont {T.}~\bibnamefont {Feng}}, \bibinfo {author} {\bibfnamefont {L.}~\bibnamefont {Lindsay}},\ and\ \bibinfo {author} {\bibfnamefont {X.}~\bibnamefont {Ruan}},\ }\bibfield  {title} {\bibinfo {title} {Four-phonon scattering significantly reduces intrinsic thermal conductivity of solids},\ }\href {https://doi.org/10.1103/PhysRevB.96.161201} {\bibfield  {journal} {\bibinfo  {journal} {Phys. Rev. B}\ }\textbf {\bibinfo {volume} {96}},\ \bibinfo {pages} {161201} (\bibinfo {year} {2017})}\BibitemShut {NoStop}%
\bibitem [{\citenamefont {Tong}\ \emph {et~al.}(2020)\citenamefont {Tong}, \citenamefont {Yang}, \citenamefont {Feng}, \citenamefont {Bao},\ and\ \citenamefont {Ruan}}]{PhysRevB.101.125416}%
  \BibitemOpen
  \bibfield  {author} {\bibinfo {author} {\bibfnamefont {Z.}~\bibnamefont {Tong}}, \bibinfo {author} {\bibfnamefont {X.}~\bibnamefont {Yang}}, \bibinfo {author} {\bibfnamefont {T.}~\bibnamefont {Feng}}, \bibinfo {author} {\bibfnamefont {H.}~\bibnamefont {Bao}},\ and\ \bibinfo {author} {\bibfnamefont {X.}~\bibnamefont {Ruan}},\ }\bibfield  {title} {\bibinfo {title} {First-principles predictions of temperature-dependent infrared dielectric function of polar materials by including four-phonon scattering and phonon frequency shift},\ }\href {https://doi.org/10.1103/PhysRevB.101.125416} {\bibfield  {journal} {\bibinfo  {journal} {Phys. Rev. B}\ }\textbf {\bibinfo {volume} {101}},\ \bibinfo {pages} {125416} (\bibinfo {year} {2020})}\BibitemShut {NoStop}%
\bibitem [{\citenamefont {Lepori}\ \emph {et~al.}(2023)\citenamefont {Lepori}, \citenamefont {Trombettoni}, \citenamefont {Giuliano}, \citenamefont {Kombe}, \citenamefont {Malo}, \citenamefont {Daley}, \citenamefont {Smerzi},\ and\ \citenamefont {Chiofalo}}]{Lepori_2023}%
  \BibitemOpen
  \bibfield  {author} {\bibinfo {author} {\bibfnamefont {L.}~\bibnamefont {Lepori}}, \bibinfo {author} {\bibfnamefont {A.}~\bibnamefont {Trombettoni}}, \bibinfo {author} {\bibfnamefont {D.}~\bibnamefont {Giuliano}}, \bibinfo {author} {\bibfnamefont {J.}~\bibnamefont {Kombe}}, \bibinfo {author} {\bibfnamefont {J.~Y.}\ \bibnamefont {Malo}}, \bibinfo {author} {\bibfnamefont {A.~J.}\ \bibnamefont {Daley}}, \bibinfo {author} {\bibfnamefont {A.}~\bibnamefont {Smerzi}},\ and\ \bibinfo {author} {\bibfnamefont {M.~L.}\ \bibnamefont {Chiofalo}},\ }\bibfield  {title} {\bibinfo {title} {Can multipartite entanglement be characterized by two-point connected correlation functions?},\ }\href {https://doi.org/10.1088/1751-8121/acdd36} {\bibfield  {journal} {\bibinfo  {journal} {Journal of Physics A: Mathematical and Theoretical}\ }\textbf {\bibinfo {volume} {56}},\ \bibinfo {pages} {305302} (\bibinfo {year} {2023})}\BibitemShut {NoStop}%
\bibitem [{\citenamefont {Wang}\ \emph {et~al.}(2022)\citenamefont {Wang}, \citenamefont {Guo},\ and\ \citenamefont {Shan}}]{e24101467}%
  \BibitemOpen
  \bibfield  {author} {\bibinfo {author} {\bibfnamefont {J.}~\bibnamefont {Wang}}, \bibinfo {author} {\bibfnamefont {G.}~\bibnamefont {Guo}},\ and\ \bibinfo {author} {\bibfnamefont {Z.}~\bibnamefont {Shan}},\ }\bibfield  {title} {\bibinfo {title} {Sok: Benchmarking the performance of a quantum computer},\ }\bibfield  {journal} {\bibinfo  {journal} {Entropy}\ }\textbf {\bibinfo {volume} {24}},\ \href {https://doi.org/10.3390/e24101467} {10.3390/e24101467} (\bibinfo {year} {2022})\BibitemShut {NoStop}%
\bibitem [{\citenamefont {Zhang}\ and\ \citenamefont {Murakami}(2022)}]{PhysRevResearch.4.L012024}%
  \BibitemOpen
  \bibfield  {author} {\bibinfo {author} {\bibfnamefont {T.}~\bibnamefont {Zhang}}\ and\ \bibinfo {author} {\bibfnamefont {S.}~\bibnamefont {Murakami}},\ }\bibfield  {title} {\bibinfo {title} {Chiral phonons and pseudoangular momentum in nonsymmorphic systems},\ }\href {https://doi.org/10.1103/PhysRevResearch.4.L012024} {\bibfield  {journal} {\bibinfo  {journal} {Phys. Rev. Res.}\ }\textbf {\bibinfo {volume} {4}},\ \bibinfo {pages} {L012024} (\bibinfo {year} {2022})}\BibitemShut {NoStop}%
\bibitem [{\citenamefont {Jin}\ \emph {et~al.}(2022)\citenamefont {Jin}, \citenamefont {Hu}, \citenamefont {Liu}, \citenamefont {Li}, \citenamefont {Zhang}, \citenamefont {Iida}, \citenamefont {Kamazawa}, \citenamefont {Kolesnikov}, \citenamefont {Stone}, \citenamefont {Zhang}, \citenamefont {Chen}, \citenamefont {Wang}, \citenamefont {Zaliznyak}, \citenamefont {Tranquada}, \citenamefont {Fang},\ and\ \citenamefont {Li}}]{PhysRevB.106.224304}%
  \BibitemOpen
  \bibfield  {author} {\bibinfo {author} {\bibfnamefont {Z.}~\bibnamefont {Jin}}, \bibinfo {author} {\bibfnamefont {B.}~\bibnamefont {Hu}}, \bibinfo {author} {\bibfnamefont {Y.}~\bibnamefont {Liu}}, \bibinfo {author} {\bibfnamefont {Y.}~\bibnamefont {Li}}, \bibinfo {author} {\bibfnamefont {T.}~\bibnamefont {Zhang}}, \bibinfo {author} {\bibfnamefont {K.}~\bibnamefont {Iida}}, \bibinfo {author} {\bibfnamefont {K.}~\bibnamefont {Kamazawa}}, \bibinfo {author} {\bibfnamefont {A.~I.}\ \bibnamefont {Kolesnikov}}, \bibinfo {author} {\bibfnamefont {M.~B.}\ \bibnamefont {Stone}}, \bibinfo {author} {\bibfnamefont {X.}~\bibnamefont {Zhang}}, \bibinfo {author} {\bibfnamefont {H.}~\bibnamefont {Chen}}, \bibinfo {author} {\bibfnamefont {Y.}~\bibnamefont {Wang}}, \bibinfo {author} {\bibfnamefont {I.~A.}\ \bibnamefont {Zaliznyak}}, \bibinfo {author} {\bibfnamefont {J.~M.}\ \bibnamefont {Tranquada}}, \bibinfo {author} {\bibfnamefont {C.}~\bibnamefont {Fang}},\ and\ \bibinfo {author} {\bibfnamefont {Y.}~\bibnamefont {Li}},\
  }\bibfield  {title} {\bibinfo {title} {Chern numbers of topological phonon band crossing determined with inelastic neutron scattering},\ }\href {https://doi.org/10.1103/PhysRevB.106.224304} {\bibfield  {journal} {\bibinfo  {journal} {Phys. Rev. B}\ }\textbf {\bibinfo {volume} {106}},\ \bibinfo {pages} {224304} (\bibinfo {year} {2022})}\BibitemShut {NoStop}%
\bibitem [{\citenamefont {Murta}\ \emph {et~al.}(2020)\citenamefont {Murta}, \citenamefont {Catarina},\ and\ \citenamefont {Fernández-Rossier}}]{Murta_2020}%
  \BibitemOpen
  \bibfield  {author} {\bibinfo {author} {\bibfnamefont {B.}~\bibnamefont {Murta}}, \bibinfo {author} {\bibfnamefont {G.}~\bibnamefont {Catarina}},\ and\ \bibinfo {author} {\bibfnamefont {J.}~\bibnamefont {Fernández-Rossier}},\ }\bibfield  {title} {\bibinfo {title} {Berry phase estimation in gate-based adiabatic quantum simulation},\ }\bibfield  {journal} {\bibinfo  {journal} {Physical Review A}\ }\textbf {\bibinfo {volume} {101}},\ \href {https://doi.org/10.1103/physreva.101.020302} {10.1103/physreva.101.020302} (\bibinfo {year} {2020})\BibitemShut {NoStop}%
\end{thebibliography}%
\end{document}